\documentclass[preprint,longabstract]{aastex}
\usepackage{graphicx}
\begin{document}

\title{The 2004--2006 outburst and environment of V1647 Ori}

\author{J. A. Acosta--Pulido}
\affil{Instituto de Astrof\'{\i}sica de Canarias, E-38200, La Laguna, Tenerife, 
Canary Islands, Spain}
\email{jap@iac.es}

\author{M. Kun, P.~\'Abrah\'am, \'A.~K\'osp\'al, Sz.~Csizmadia}
\affil{Konkoly Observatory, H-1525 Budapest, P.O. Box 67, Hungary}

\author{L. L. Kiss\altaffilmark{1}}
\affil{School of Physics, University of Sydney, Australia} 

\author{A. Mo\'or, L. Szabados, J. M. Benk\H{o}}
\affil{Konkoly Observatory, H-1525 Budapest, P.O. Box 67, Hungary} 

\author{R. Barrena Delgado, M. Charcos--Llorens\altaffilmark{2}}
\affil{Instituto de Astrof\'{\i}sica de Canarias, E-38200, La Laguna, Tenerife, 
Canary Islands, Spain} 

\author{M. Eredics}
\affil{Konkoly Observatory, H-1525 Budapest, P.O. Box 67, Hungary} 

\author{Z. T. Kiss}
\affil{Baja Astronomical Observatory, P.O. Box 766, H-6500 Baja, Hungary} 

\author{A. Manchado\altaffilmark{3}}
\affil{Instituto de Astrof\'{\i}sica de Canarias, E-38200, La Laguna, Tenerife, 
Canary Islands, Spain} 

\author{M. R\'acz}
\affil{Konkoly Observatory, H-1525 Budapest, P.O. Box 67, Hungary} 

\author{C. Ramos Almeida}
\affil{Instituto de Astrof\'{\i}sica de Canarias, E-38200, La Laguna, Tenerife, 
Canary Islands, Spain} 

\author{P. Sz\'ekely}
\affil{Dept. of Experimental Physics and Astronomical Observatory, 
 University of Szeged, Hungary} 

\and

\author{M. J. Vidal--N\'u\~{n}ez\altaffilmark{4}}
\affil{Instituto de Astrof\'{\i}sica de Canarias, E-38200, La Laguna, Tenerife, 
Canary Islands, Spain} 

\altaffiltext{1}{On leave from University of Szeged, Hungary}
\altaffiltext{2}{Dept. of Astronomy, Univ. of Florida, USA}
\altaffiltext{3}{Consejo Superior de Investigaciones Cientificas, Spain}
\altaffiltext{4}{Instituto de Astrof\'{\i}sica de Andaluc\'{\i}a, Granada, Spain}

\date{Received \today / Accepted}
                                                                                
\begin{abstract} We studied  the brightness and spectral evolution
of the young eruptive star V1647~Ori during its recent outburst in
the period 2004 February -- 2006 Sep. 
We performed a photometric follow--up in the bands {\it V\/},  
{\it R\/}$_\mathrm{C}$, {\it I\/}$_\mathrm{C}$,  
{\it J\/},  {\it H\/}, {\it K\/}$_s$ as well as visible and 
near-IR spectroscopy. 
The main results derived from combining our data with those published 
by other authors are as follows: The brightness
of V1647~Ori stayed more than  4\,mag above the pre-outburst level
until 2005 October when it started a  rapid fading. In the high
state we found a periodic component in the optical light curves
with a period of 56 days.  The delay between variations of the 
star and variations in the brightness of clump of nearby nebulosity 
corresponds to an angle of $61\degr\pm14\degr$ between the axis  
of the nebula and the line of sight.  
The overall appearance of the  infrared and
optical spectra did not change in the period  March~2004 --
March~2005, though a steady decrease of HI emission line  fluxes
could be observed. 
In 2006 May, in the quiescent phase, the HeI 1.083 $\mu$m line was
observed in  emission, contrary to its deep blueshifted absorption observed
during the outburst.  The $J-H$ and $H-K_s$ color maps  of the
infrared nebula reveal an envelope around the star whose largest 
extension is about 18\arcsec (0.03\,pc).  The color distribution
of the infrared  nebula suggests reddening of the scattered light
inside a thick circumstellar  disk. Comparison of the {\it K\/}$_s$  
and H$\alpha$ images of McNeil's Nebula, the conical nebulosity 
illuminated by V1647~Ori, shows that HH\,22A, the Spitzer infrared source 
and the bright clump~C
of the nebula may be unrelated objects.  We show that the observed
properties of V1647~Ori could be interpreted  in the framework of
the thermal instability models of Bell et al. (1995).
V1647~Ori might belong to a new class of young eruptive
stars, defined by relatively short  timescales, recurrent
outbursts, modest increase in bolometric luminosity and accretion
rate, and an evolutionary state earlier than that of typical
EXors.  

\end{abstract}


\section{Introduction}
\label{Sect_1}

Eruptive young stellar objects form a small, but spectacular class of  pre-main
sequence stars.   Traditionally they are divided into two groups.  FU
Orionis-type stars (FUors) are characterized by an initial brightening of
$\sim$5\,mag during several months or years, followed by a fading phase of up
to several decades or a century. Their spectral type is F--G giant according to
the optical spectrum, and K--M giant/supergiant at near-infrared wavelengths.
The second group, called EX Lupi-type stars (EXors), belongs to the T\,Tauri
class. Their recurrent outbursts are relatively short, lasting from some weeks
to months,  and the time between the eruptions ranges from months to years.
Their spectral type is K or M dwarf.

The recent outburst of the young star V1647~Ori offered a rare opportunity to 
study  phenomena accompanying the eruption of a low-mass pre-main sequence star. 
Following a 3-month brightening of ${\Delta}I_{C} \sim 4.5$\,mag 
\citep{Briceno04},
V1647~Ori reached peak light in January 2004, illuminating also a conical reflection 
nebulosity called McNeil's Nebula \citep{McNeil}. The object stayed in  
high state for about 2 years, then started a rapid fading in October 
2005 \citep{Kospal05}. 
V1647~Ori exhibited already a similar eruption in 1966--67 \citep{Aspin06}.

During the 2004--2005 outburst V1647~Ori was intensively studied from many aspects.
The brightness evolution at optical and near-infrared (NIR) wavelengths was monitored 
by \citet{Briceno04}, \citet{RA}, \citet{Walter04}, \citet{McGehee}, and \citet{Ojha05,Ojha06}. The
optical/infrared spectrum, and its changes were studied by \citet{VCS}, \citet{Walter04},
\citet{ARS}, \citet{Rettig}, \citet{Ojha06}, and \citet{Gibb06}. The structure of the 
circumstellar matter
and the physics of the outburst were investigated by \citet{McGehee},
\citet{Muzerolle04}, and \citet{Rettig}. 
The X-ray properties of the star were studied by 
\citet{Grosso}, \citet{Kastner}, and \citet{Kastner06}.
These works produced a wealth of very important information on the eruption,
but a general picture is still to be worked out.
Nevertheless, all authors seem to agree that: 
V1647~Ori is a deeply embedded low-mass pre-main sequence object; it is surrounded 
by a disk; the outburst was caused by increased
accretion; and that the eruption was accompanied by 
strong stellar wind. It is still an open question
whether V1647~Ori belongs to the FUor or the EXor class, and
whether the standard outburst models, developed for FU Orionis events, could be
applied for this case (or how should they be modified to make them applicable).

In this paper we present and analyse the results of our monitoring programme
on V1647~Ori. The data, which
include optical ({\it V\/}, {\it R\/}$_\mathrm{C}$, 
{\it I\/}$_\mathrm{C}$) and near-infrared ({\it J\/}, {\it H\/}, {\it K\/}$_s$) imaging,
and intermediate-resolution optical and near-infrared spectroscopic observations, 
were obtained between February 2004 and May 2006, covering the whole outburst 
period. Following a description of the observations and data reduction 
(Sect.~\ref{sc:obs}), in Sect.~3 we present results on the photometric evolution
of V1647~Ori. Sects. 4 and 5 are devoted to the spectroscopic
variability and the morphology of the nebula, respectively. In Sect.~6
we discuss the physics of the outburst, and comment on the
FUor-or-EXor question. 
Our results are briefly summarized in Sect.~\ref{sc:conclu}.

\section{Observations and data reduction}
\label{sc:obs}

\subsection{Optical imaging and photometry}

Optical observations of V1647\,Ori were obtained on 45 nights in the period
2004 February 12 -- 2006 February 1. The measurements were performed in
two observatories with three telescopes: the IAC--80 telescope of the
Teide Observatory (Spain), and the 1\,m Ritchey-Chr\'etien-Coud\'e (RCC) 
and 60/90/180\,cm Schmidt telescopes of the Konkoly Observatory
(Hungary).

The IAC--80 telescope was equipped with a 1024$\times$1024 Thomson CCD
with a scale of 0\farcs4325/pixel, and a field of view of
7\farcm4$\times$7\farcm4. 
For the broad-band {\it VRI} observations we used the IAC\#72, IAC\#71,
and IAC\#70 filters\footnote{The 
transmission curves are available at 
http://www.iac.es/telescopes/tcs/filtros-eng.htm}. 
Typically three frames were
taken with exposure times of 120--300\,s/frame each
night. Additionally, on the night of 2004 January 11 deep H$\alpha$ and 
S[II] images were obtained using the narrow band filters IAC\#17 and
IAC\#68. Four exposures of 1200~s were taken through each filter.  

At Konkoly Observatory, the 1\,m RCC telescope was equipped with a
Princeton Instruments VersArray 1300B camera. The back-illuminated
1340$\times$1300 CCD has an image scale of 0\farcs306/pixel and a field
of view of 6\farcm8$\times$6\farcm6. The Schmidt telescope
was equipped with a 1536$\times$1024 Kodak Photometrics AT\,200 CCD
camera. The image scale was 1\farcs03/pixel, providing a
field of view of $24\arcmin\times17\arcmin$. On both telescopes,
the utilized {\it V(RI)}$_{\mathrm{C}}$ filters matched closely
the standard Johnson-Cousins system \citep{Hamilton05}. 
For each filter, 3--10 frames were taken on 
each night, and the integration times varied between 120 and 600\,s/frame.

\begin{figure*}
\centerline{\includegraphics[width=5.5cm]{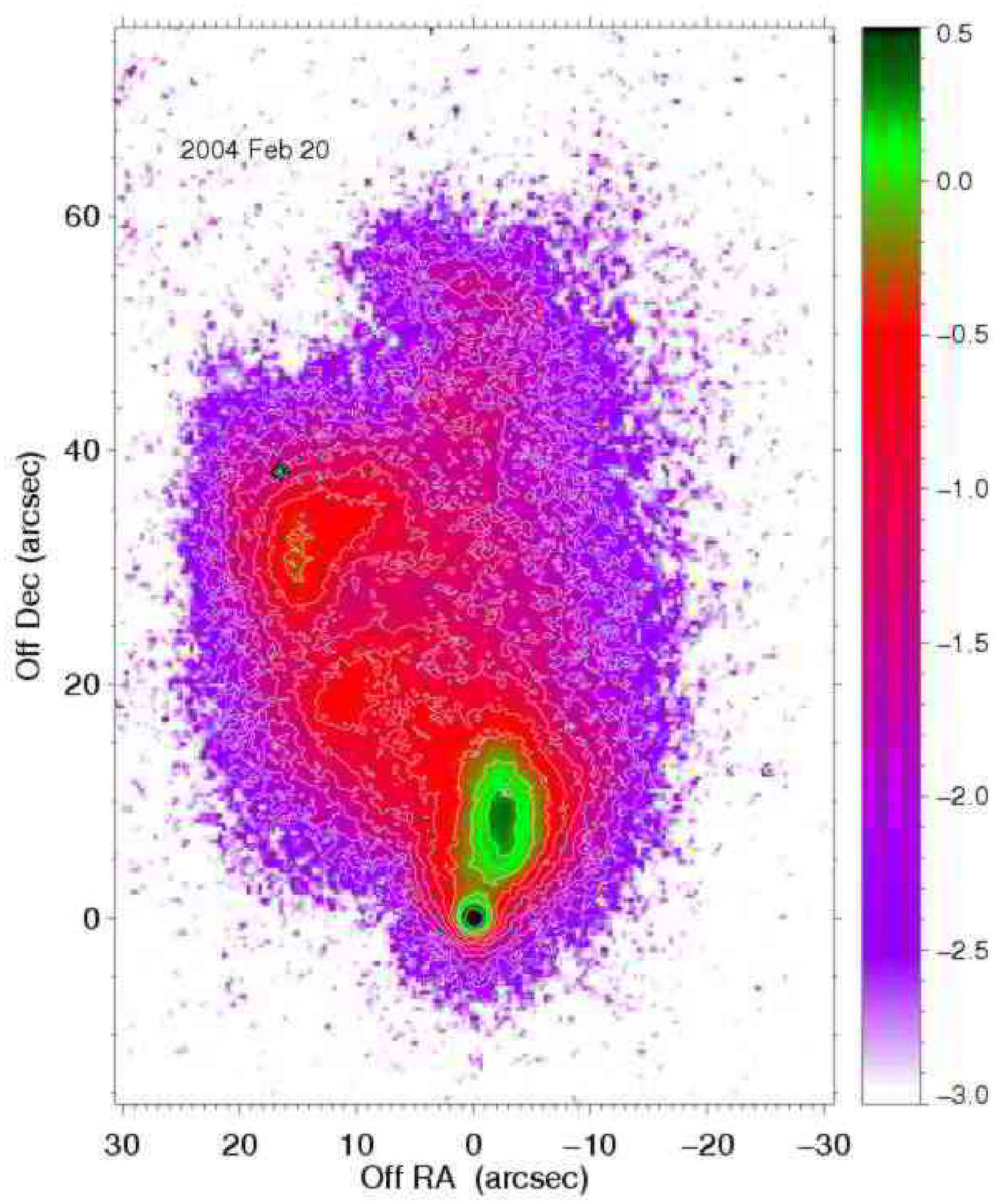}
\hspace{5mm}\includegraphics[width=5.5cm]{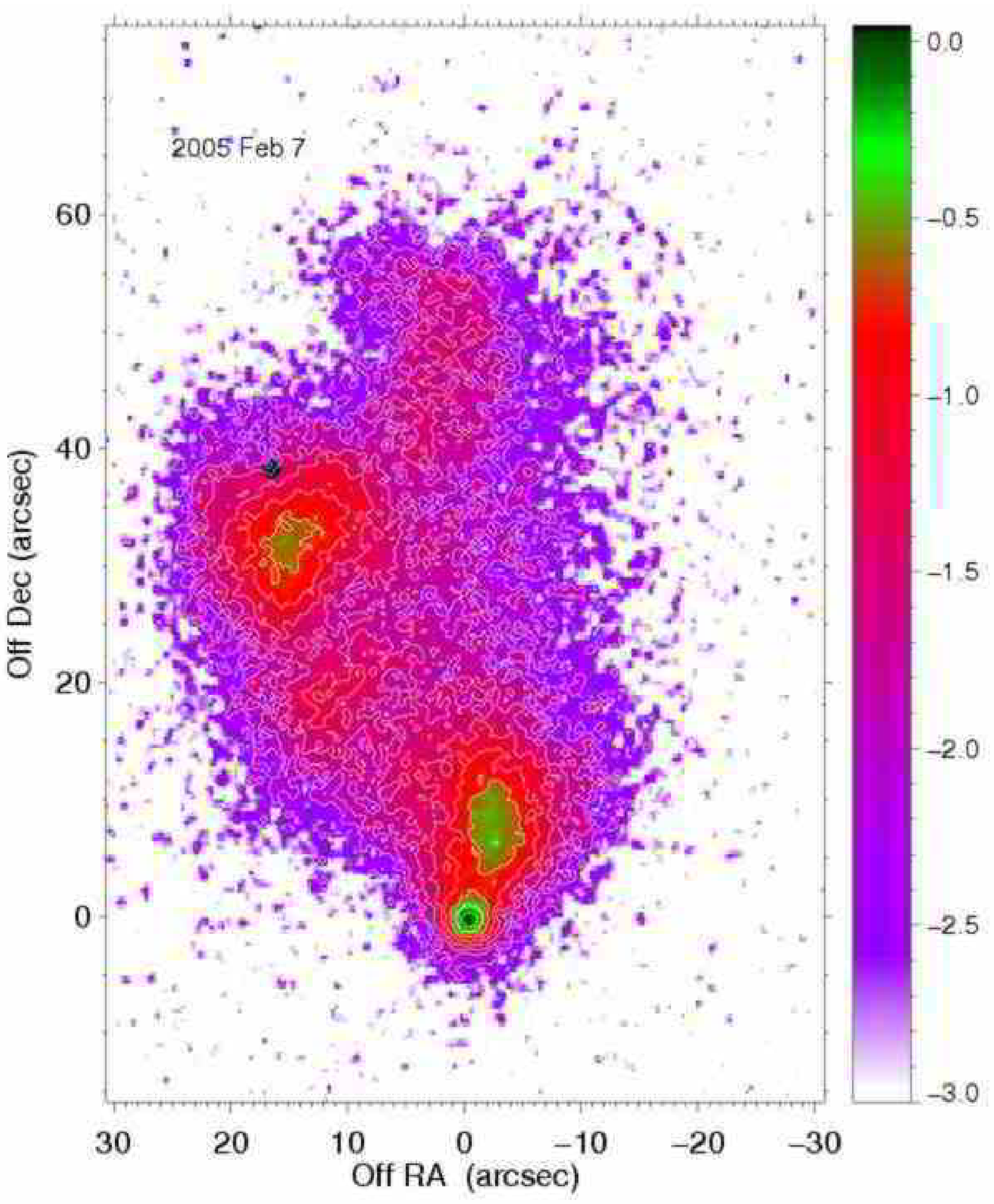}
\hspace{5mm}\includegraphics[width=5.5cm]{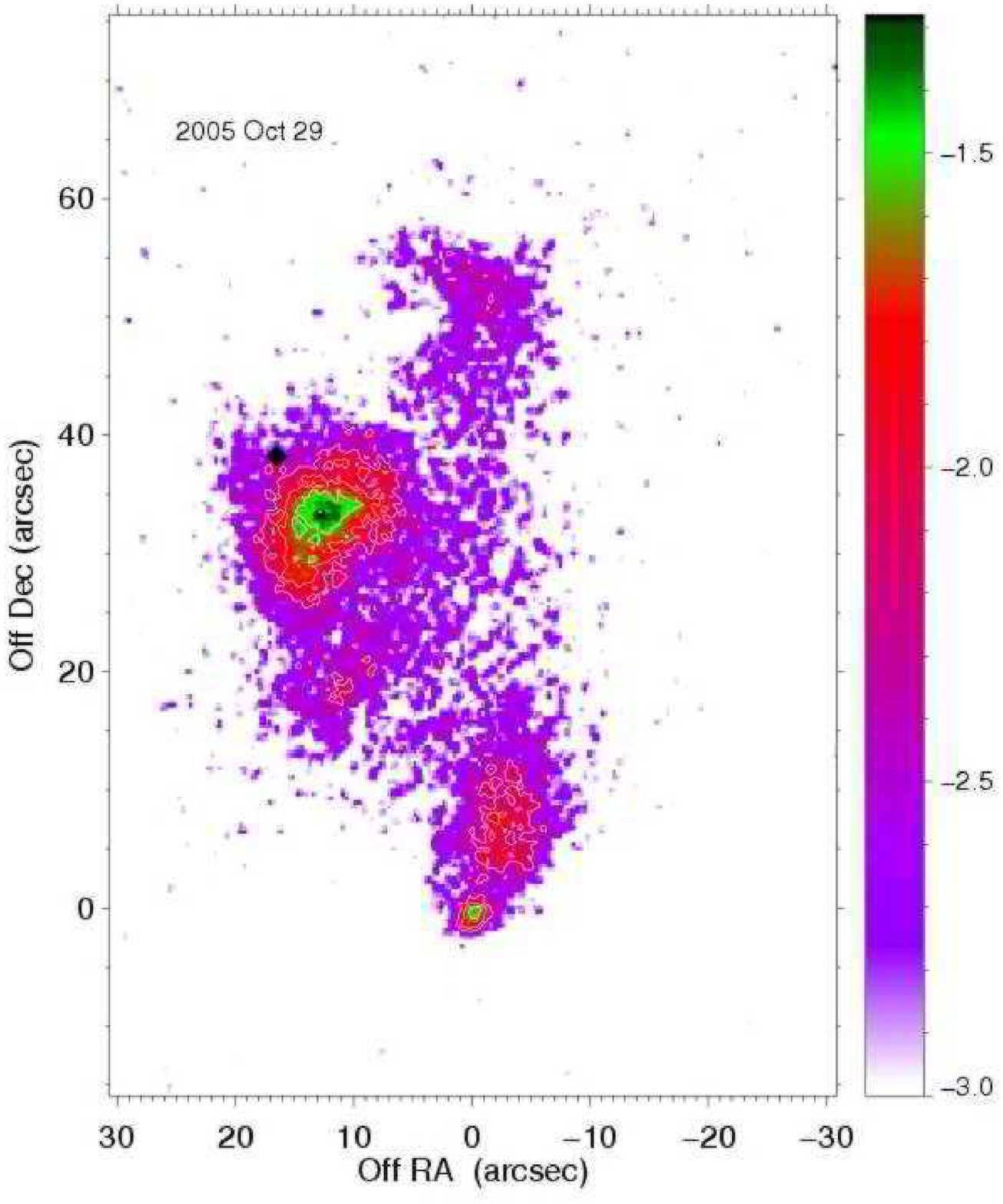}}
\caption{Examples of R-band optical images of V1647~Ori taken with the 1m RCC
telescope of the Konkoly Observatory at three different phases of the outburst.
North is at the top and east to the left. V1647\,Ori is located at the southern
apex of the nebula. The color table has been scaled to represent the
same brightness in all three images.}
\label{fig:opt_examp}
\end{figure*} 

\paragraph{Data reduction and photometric calibration.}

Raw frames were dark-subtracted and flat-fielded using the `imred'
package within the IRAF\footnote{IRAF is
distributed by the National Optical Astronomy Observatories, which
are operated by the Association of Universities for the Research in
Astronomy, Inc., under cooperative agreement with the National Science
Foundation. http://iraf.noao.edu/} environment. Dome flat-field images were 
taken each night, and sometimes sky flats were also available. In some
cases, the consecutive frames taken through the same filter were 
co-added in order to increase the signal to noise. 
As an example, Fig.~\ref{fig:opt_examp} shows R-band optical images of
V1647~Ori taken with the 1m RCC telescope of the Konkoly Observatory at
three different phases of the outburst. 
The narrow-band H$\alpha$
and S[II] images were bias-subtracted, flat-fielded and combined
per filter. Then a scaled R-band image, assumed to be a proper
continuum, was subtracted from them in order to show the pure 
emission line map.

The instrumental {\it V(RI)}$_{\mathrm{C}}$ magnitudes of V1647~Ori 
were determined by
PSF-photometry using the `daophot' package in IRAF \citep{Stetson87}. 
The field of view
of our telescopes allowed us to include several field stars to define
the point spread function of the images. The use of PSF--photometry 
was important to exclude the contribution of the nebula, which becomes
dominant especially in the V-band. 
In order to assure the consistency of the results obtained with the 
three different telescopes we treated our data in a homogeneous manner:
the position and extension of the sky region near V1647~Ori were the same 
in all images, and the preliminary aperture photometry, used for scaling
the PSF magnitudes was obtained using equally narrow apertures 
(1.5 arcsec) in each image.  

\begin{figure*}
\centerline{\includegraphics[width=14cm]{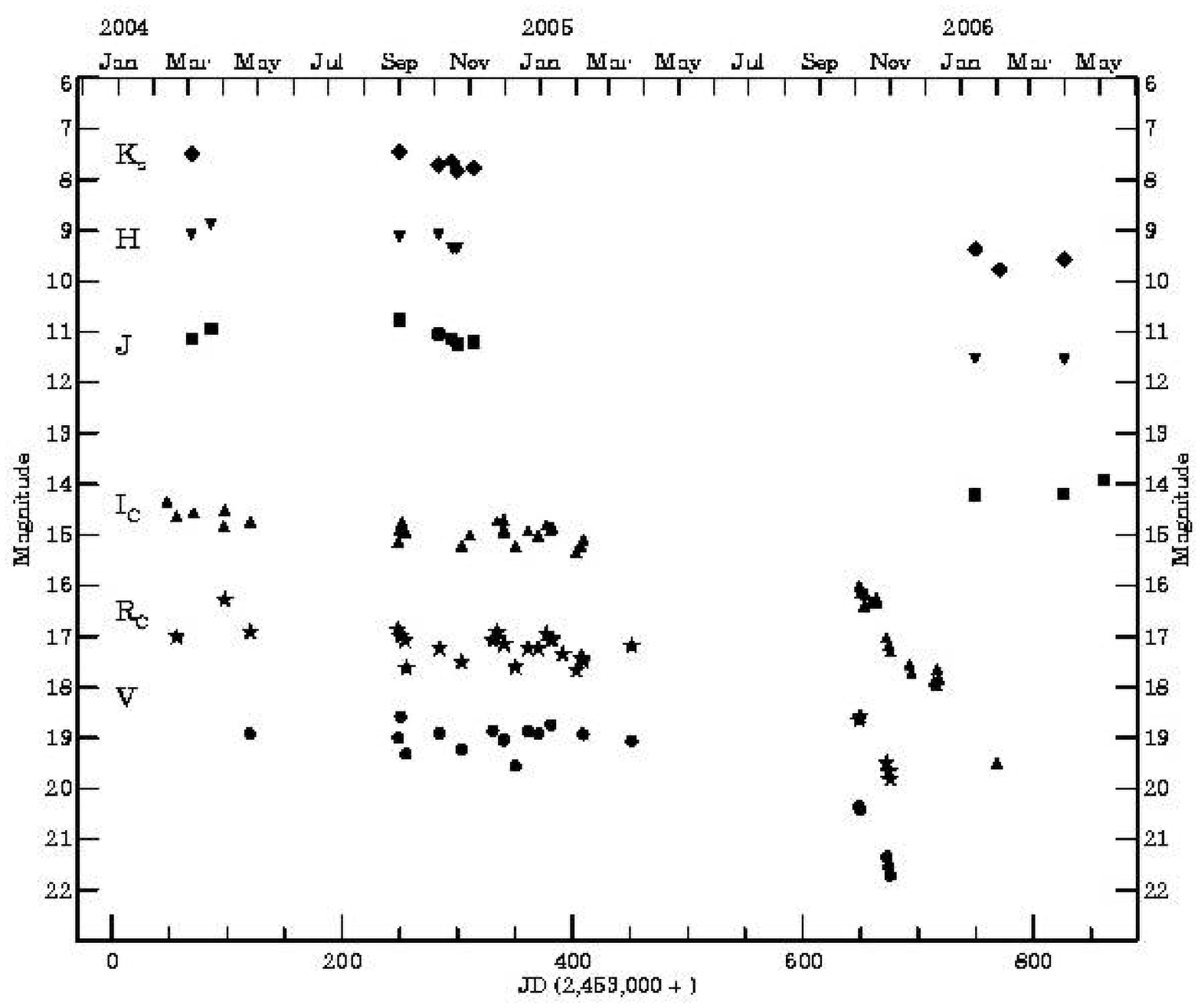}}
\caption{Light curves of V1647 Ori in the {\it V\/(dots)}, 
{\it R\/}$_\mathrm{C}$ {\it (asterisks)}, {\it I\/}$_\mathrm{C}$
{\it (upward triangles)}, {\it J\/(squares)}, {\it H\/(downward triangles)},
and {\it K\/}$_s$ {\it (diamonds)} bands between 2004 February 16 
and 2006 May 5 , based on the data presented in Tables~\ref{tab:opdata} and
\ref{ta:irdata}. 
The uncertainties are comparable with symbol sizes.}
\label{fi:allbphot}
\end{figure*}

For photometric calibration we used the secondary standards from the list 
of \citet{Semkov06}.
Differential photometry was performed by
computing the instrumental magnitude differences between V1647\,Ori
and each comparison star.
The differential magnitudes were transformed
into the standard photometric system as follows.
For the comparison stars,
we searched for a relationship between $\Delta m$, the differences 
between instrumental and absolute magnitudes in the {\it V\/}, 
{\it R\/}$_\mathrm{C}$ or {\it I\/}$_\mathrm{C}$ band and the color indices
of the stars, similarly as described in \citet{Serra99}.
We note that the color indices of the comparison stars covered a large
enough range to bracket the color indices of V1647\,Ori. The
data points were fitted with a linear relationship, except a few
low-quality nights, when due to large scatter, all $\Delta m$ values
were averaged. For a given telescope, the derived color terms
(the slope of the fit) were
consistent from night to night within the formal uncertainties
provided by the fitting procedure. Color terms computed from the 1\,m RCC
observations also agreed with values determined with high precision from
observations of NGC\,7790 with the same instrument.
For calibration of the V1647\,Ori data we applied the color terms
determined for the corresponding night. Results of the optical photometry of
V1647\,Ori are presented in Tab.~\ref{tab:opdata} and plotted in
Fig.~\ref{fi:allbphot}. The uncertainties given in Tab.~\ref{tab:opdata} were
computed as the quadratic sum of the formal errors of the instrumental
magnitudes provided by IRAF and the uncertainties of the standard
transformation. In those cases when the frames were evaluated
separately (rather than co-added), the standard deviation of the
magnitudes from each frame was also computed, and the maximum of the
two types of errors was adopted.

\begin{deluxetable}{lrcccl}
\tablecolumns{6}
\tablewidth{0pt}
\tablecaption{\label{tab:opdata} Optical photometry of V1647~Ori}
\tablehead{
\colhead{Date} &  \colhead{JD\tablenotemark{a}}  &  \colhead{{\it V\/}} 
& \colhead{{\it R\/}$_\mathrm{C}$} & \colhead{{\it I\/}$_\mathrm{C}$} & 
\colhead{Telescope}}
\startdata
2004\,02\,12 & 048.3 &  \nodata      &   \nodata     & 14.33\,(0.04) & Schmidt \\
2004\,02\,20 & 056.3 &  \nodata      & 17.01\,(0.08) & 14.62\,(0.04) & RCC     \\
2004\,03\,06 & 071.3 &  \nodata      &  \nodata      & 14.56\,(0.04) & RCC     \\
2004\,04\,01 & 097.2 &  \nodata      &  \nodata      & 14.83\,(0.08) & Schmidt \\
2004\,04\,02 & 098.3 &  \nodata      & 16.28\,(0.08) & 14.51\,(0.06) & Schmidt \\
2004\,04\,24 & 120.3 & 18.92\,(0.06) & 16.91\,(0.04) & 14.73\,(0.07) & IAC-80  \\
2004\,08\,31 & 248.6 & 19.00\,(0.12) & 16.86\,(0.10) & 15.13\,(0.15) & RCC     \\
2004\,09\,01 & 249.6 &  \nodata      &   \nodata     & 14.89\,(0.08) & IAC-80  \\
2004\,09\,02 & 250.6 & 19.58\,(0.08) & 17.00\,(0.08) &   \nodata     & IAC-80  \\
2004\,09\,03 & 251.6 &  \nodata      &    \nodata    & 14.74\,(0.04) & RCC     \\
2004\,09\,06 & 254.6 &  \nodata      & 17.08\,(0.03) & 14.90\,(0.04) & RCC     \\
2004\,09\,07 & 255.6 & 19.32\,(0.06) & 17.62\,(0.08) & 14.95\,(0.03) & RCC     \\
2004\,10\,06 & 284.7 & 18.91\,(0.10) & 17.24\,(0.06) & \nodata       & IAC-80  \\
2004\,10\,25 & 303.6 & 19.23\,(0.10) & 17.50\,(0.05) & 15.21\,(0.06) & IAC-80  \\
2004\,11\,01 & 310.8 &  \nodata      &  \nodata      & 14.99\,(0.04) & IAC-80  \\
2004\,11\,21 & 330.7 & 18.86\,(0.08) & 17.08\,(0.08) &  \nodata      & IAC-80  \\
2004\,11\,24 & 334.5 &               & 16.91\,(0.07) & 14.72\,(0.03) & RCC     \\
2004\,11\,30 & 340.4 & 19.04\,(0.06) & 17.16\,(0.04) & 14.87\,(0.04) & RCC     \\
2004\,12\,11 & 350.5 & 19.56\,(0.08) & 17.60\,(0.05) & 15.23\,(0.06) & RCC     \\
2004\,12\,22 & 361.5 & 18.87\,(0.11) & 17.24\,(0.08) & 14.91\,(0.05) & RCC     \\
2004\,12\,30 & 370.3 & 18.92\,(0.05) & 17.24\,(0.06) & 15.01\,(0.05) & RCC     \\
2005\,01\,06 & 377.3 & \nodata       & 16.95\,(0.06) & 14.79\,(0.07) & RCC     \\
2005\,01\,10 & 381.3 & 18.74\,(0.06) & 17.08\,(0.08) & 14.91\,(0.10) & RCC     \\
2005\,01\,11 & 382.3 & \nodata       & 17.04\,(0.07) & 14.84\,(0.07) & RCC     \\
2005\,01\,20 & 391.3 & \nodata       & 17.35\,(0.06) &  \nodata      & IAC-80  \\
2005\,02\,01 & 403.3 & \nodata       & 17.67\,(0.04) & 15.34\,(0.10) & RCC     \\
2005\,02\,05 & 407.3 & \nodata       & 17.43\,(0.09) & 15.23\,(0.05) & RCC     \\
2005\,02\,07 & 409.3 & 18.93\,(0.09) & 17.48\,(0.07) & 15.09\,(0.05) & RCC     \\
2005\,03\,21 & 451.3 & 19.06\,(0.07) & 17.18\,(0.03) &  \nodata      & IAC-80  \\
2005\,10\,04 & 648.5 & 20.35\,(0.10) & 18.65\,(0.06) & 16.00\,(0.08) & RCC     \\
2005\,10\,05 & 649.5 & 20.43\,(0.06) & 18.58\,(0.04) & 16.14\,(0.05) & RCC     \\
2005\,10\,09 & 653.6 &   \nodata     &   \nodata     & 16.39\,(0.06) & RCC     \\
2005\,10\,10 & 654.6 &   \nodata     &   \nodata     & 16.16\,(0.04) & RCC     \\
2005\,10\,15 & 659.7 &   \nodata     &   \nodata     & 16.03\,(0.07) & RCC     \\
2005\,10\,19 & 663.6 &   \nodata     &    \nodata    & 16.29\,(0.05) & RCC     \\
2005\,10\,28 & 672.6 & 21.34\,(0.16) & 19.49\,(0.08) & 17.02\,(0.08) & RCC     \\
2005\,10\,31 & 674.6 & 21.55\,(0.16) & 19.66\,(0.05) & 17.17\,(0.03) & RCC     \\
2005\,11\,01 & 675.6 & 21.75\,(0.08) & 19.83\,(0.04) & 17.30\,(0.04) & RCC     \\
2005\,11\,17 & 692.5 &  \nodata      &   \nodata     & 17.55\,(0.08) & RCC     \\
2005\,11\,19 & 694.6 & \nodata       &   \nodata     & 17.70\,(0.10) & RCC     \\
2005\,12\,08 & 713.4 & \nodata       &    \nodata    & 17.88\,(0.10) & RCC     \\
2005\,12\,10 & 715.4 & \nodata       &   \nodata     & 17.95\,(0.10) & RCC     \\
2005\,12\,11 & 716.4 & \nodata       &   \nodata     & 17.63\,(0.10) & RCC     \\
2005\,12\,13 & 718.5 &  \nodata      &    \nodata    & 17.84\,(0.10) & RCC     \\
2006\,02\,01 & 768.3 & \nodata       &    \nodata    & 19.49\,(0.10) & RCC     \\
\enddata
\tablenotetext{a}{2,453,000+} 
\end{deluxetable}

\subsection{Near-infrared imaging,  photometry and polarimetry}
\label{Sect_IR}

Near-infrared J, H and K$_s$ band observations were carried out at 10
epochs in the period 2004 March -- 2006 March.
The data were obtained using two instruments: LIRIS, installed
on the 4.2\,m William Herschel Telescope (WHT) at the Observatorio del Roque
de Los Muchachos, and CAIN-2, installed on the 1.52\,m Carlos Sanchez Telescope
(CST)  at the Teide Observatory. 
Table \ref{ta:irdata} lists the epochs of all near-infrared observations.
In addition, polarization images through a J filter were obtained during 
the night of 2004 October 28, using the instrument 
LIRIS. 

\begin{deluxetable}{lrcccl}
\tablecolumns{6}
\tablewidth{0pt}
\tablecaption{\label{ta:irdata} Near infrared photometry 
of V1647 Ori}
\tablehead{
\colhead{Date} &  \colhead{JD\tablenotemark{a}}  &  \colhead{J} 
& \colhead{{\it H\/}} & \colhead{{\it K\/}$_s$} & \colhead{Telescope}}
\startdata
2004\,03\,04 & 069.4 & 11.13\,(0.10) & 9.09\,(0.07) & 7.48\,(0.09)  & LIRIS \\
2004\,03\,21 & 086.3 & 11.00\,(0.12) & 8.70\,(0.19) & \nodata & TCS \\
2004\,09\,01 & 249.6 & 11.12\,(0.10) & 9.17\,(0.03) & 7.66\,(0.01) &  TCS \\
2004\,10\,05 & 283.7 & 11.06\,(0.07) & 9.07\,(0.14) & 7.76\,(0.05) & TCS \\
2004\,10\,17 & 295.7 & 11.10\,(0.09) & 9.09\,(0.14) & 7.62\,(0.04) & TCS \\
2004\,10\,21 & 299.8 & 11.17\,(0.08) & 9.19\,(0.05) & 7.84\,(0.08) & TCS \\
2004\,11\,04 & 313.7 & 11.20\,(0.06) & \nodata & 7.77\,(0.06) &  LIRIS \\
2006\,01\,14 & 749.6 &  14.27\,(0.09) & 11.48\,(0.08) & 9.41\,(0.19) & LIRIS \\
2006\,02\,06 & 770.8 &  \nodata & \nodata & 9.78\,(0.02) & TCS \\
2006\,03\,31 & 826.5 & 14.19\,(0.13)  & 11.55\,(0.05) & 9.57\,(0.08) & TCS \\
2006\,05\,05 & 861.4 & 13.92\,(0.08)  & \nodata & \nodata & LIRIS \\
2006\,09\,09 & 987.8 & \nodata  & 11.59\,(0.05) & 9.82\,(0.06) & LIRIS \\
\enddata
\tablenotetext{a}{2,453,000+} 
\end{deluxetable}


LIRIS is a near-infrared camera/spectrograph attached to the Cassegrain focus
of the WHT \citep{Acosta03,Manchado03}. 
The data were obtained during the commissioning period (2004 March) plus 
some nights corresponding to the guaranteed time of the instrument team. 
LIRIS is based on a Hawaii-I detector with an image scale of $0\farcs25$ 
per pixel and 
$4\farcm2\times4\farcm2$ field of view. 
Observations were always performed by taking
several dithered exposures around the position of V1647~Ori in order
to ensure  proper sky image subtraction. For each dither point we took 
several frames using short exposure time to avoid saturation, 
which were later averaged to obtain the final reduced image.  
The minimum exposure time was 1~s for individual exposures,
the total integration time per filter was about 5 min. During the 
first observing period the individual exposure times were optimized 
to detect the nebula, which nearly saturated the detector at
the position of the illuminating star.

\begin{figure}
\centerline{\includegraphics[width=8.5truecm]{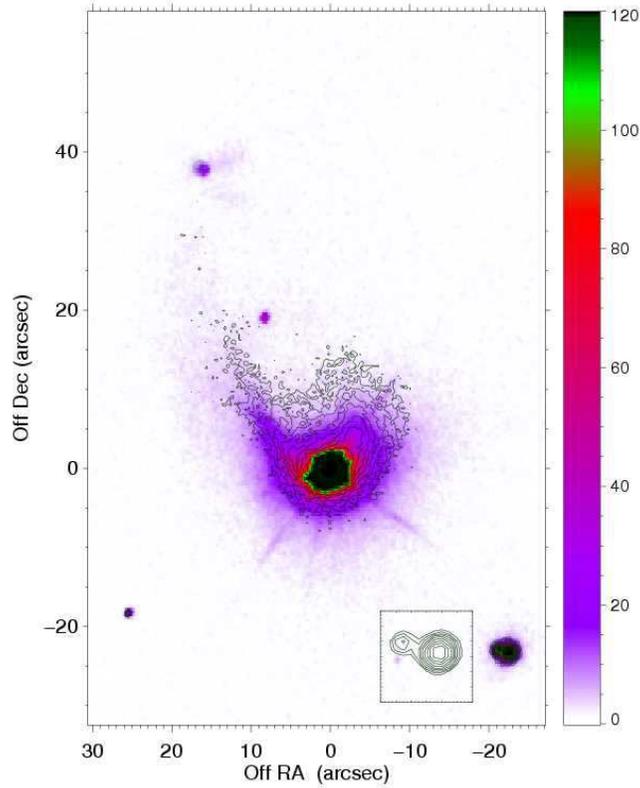}}
\caption{ Near-infrared images of McNeil's Nebula 
obtained on 2004 November 4 with LIRIS on WHT. North is at the top and east 
to the left. The color scaled image represents 
the K$_s$-band emission and the contours represent the J-band emission. 
The inset is a zoom on the 
star southwest of V1647 Ori, demonstrating that the star is a double 
system.  The open diamond at $(\Delta \alpha, \Delta \delta)
\simeq (16\arcsec,38\arcsec)$ marks the position 
of the bright infrared source detected
by \citet{Muzerolle04} in Spitzer images.}
\label{fi:lirisJoverK}
\end{figure} 

The infrared camera CAIN--2 at the 1.52\,m TCS is 
equipped with a 256$\times$256 Nicmos~3 detector, which provides a pixel 
projection of 1\arcsec \ with its wide optics configuration. 
The same dithering technique was used as in the case of LIRIS observations.
Total integration time was usually 10 min in all filters (J, H and K$_s$),
splitted in many exposures of 0.5--3~s, the shortest corresponding
to filter K$_s$ and the longest to filter J. 

\subsubsection{Data reduction and photometric calibration}

Data reduction process involved the following steps: sky subtraction,
flat-fielding and coaddition of all frames taken with
the same filter. The sky image was obtained as the median combination
of all frames, masking regions occupied by bright sources. The final image
was produced using the standard ``shift--and--add'' technique including
rejection of outlier pixels. The LIRIS images were reduced using the
package `liris$\_$ql' developed within the IRAF environment. The CAIN--2
data were also treated with IRAF tasks. 
The {\it J\/} and {\it K\/}$_s$ images obtained on 2004 November 4 can be 
seen in Fig.~\ref{fi:lirisJoverK}.

The instrumental magnitudes of the central star were extracted using 
aperture photometry since the contribution of the nebula in the 
near-infrared becomes negligible compared to it. Based on the measured
radial profiles we estimate that the nebula contribute to the central
star photometry at most 15\% in the
J band, but less than 5\% in the Ks band.  
Due to the high brightness of V1647~Ori some LIRIS images 
entered the non--linear regime at the position of the
star. In these cases we estimated the flux of the star
by fitting a PSF model to the non-saturated parts of the stellar profile 
using the {\it daophot} package in IRAF.   

For the photometric calibration we used the 2MASS catalogue \citep{Cutri03}.
In general, five or six 2MASS stars were found within the sky region
covered  by our images.
We have determined the offset between the instrumental and the 
calibrated 2MASS magnitudes by averaging all available stars, after removal 
of deviant sources (identified as likely variable stars by
\citealp{Semkov06}).
Results of the 
near-infrared photometry for V1647~Ori are listed in Tab.~\ref{ta:irdata}, 
and the light curves are shown in Fig.~\ref{fi:allbphot}.
The standard deviation is used as an estimate of the
measurement error and reported in Tab.~\ref{ta:irdata}. 
We cannot determine a reliable color term correction using our reduced
number of reference stars. 
The 2MASS star J054611162-0006279  was always  
included as a comparison star, since it shows calibration offsets
close to the average in all occasions.   
We note that this source was   
the only comparison star used by \citet{Walter04} for the
photometric calibration of their near infrared data.
The colors of this star are very similar to those of
V1647~Ori which reduces the importance of color terms when determining
the photometrical calibration.

\subsubsection{Polarimetry}

The polarization images were obtained thanks to a double Wollaston prism 
(WeDoWo) \citep{Oliva97}, located in the LIRIS grism wheel. This device
produces four different images on the detector 
from a rectangular field of $4\times 1$~arcmin$^2$. Each image corresponds
to a different linear polarization angle, namely 0, 90, 45 and 135 degrees. 
After distorsion correction, proper alignment and combination of these images, 
the degree and angle of linear polarization can be determined. 
For V1647~Ori and its nebula we obtained an exposure of about 1800~s, divided
in individual frames of 15~s. 
A small dither pattern of 3 points was performed around the source, 
interleaved with measurements at an offset position in order to determine
a proper sky emission, given the extension of the source. 
The data reduction process is very similar to that followed in normal imaging
mode, the `liris\_ql' package was used.   
Images of the linear polarization degree will be presented and discussed 
in Section \ref{sc:polariza}.

\subsection{Near-infrared spectroscopy}

Near-infrared spectrograms of V1647~Ori were obtained using LIRIS 
at five different epochs, four of them during the first year 
of the outburst. Five spectra cover the wavelength interval of 
0.9--1.4$\mu$m ({\it ZJ\/} bands) and three others cover  the 
1.4--2.4$\mu$m ({\it HK\/}) wavelength band.
The log of the observations is presented in Tab.~\ref{Tab_sp}.

\begin{deluxetable}{llccrcrc}
\tablewidth{0pt}
\tablecolumns{8}
\tablecaption{Journal of Near-infrared spectroscopy
\label{Tab_sp}}
\tablehead{
\colhead{Date} & \colhead{JD\tablenotemark{a}} & \colhead{Slit} & \colhead{Band} &
\colhead{t$_{exp}$} & \colhead{Band} & \colhead{t$_{exp}$} & \colhead{Airmass}  \\
\colhead{}  & \colhead{}  & \colhead{} & \colhead{} & \colhead{(s)}  & \colhead{}  &  \colhead{(s)} }
\startdata
2004 03 08 & 073.4 & 1           & ZJ & 400 & HK & 12 & 1.3 \\
2004 11 04 & 313.7 & $0\farcs75$ & ZJ & 800 & HK & 432 & 1.2 \\
2005 01 24 & 395.4 & $0\farcs75$ & ZJ & 200 & \nodata & \nodata & 1.2\\
2005 03 25 & 455.4 & 1           & ZJ & 100 & \nodata & \nodata & 2.0 \\
2006 05 05 & 861.4 & 1           & ZJ & 1200 & HK & 400 & 4.1 \\
2006 09 09 & 987.7 & $0\farcs75$ & ZJ & 1200 & HK & 900 & 1.4 \\
\enddata
\tablenotetext{a}{2,453,000+}
\end{deluxetable}

Observations were performed following an ABBA telescope nodding
pattern both in the ZJ and HK bands. 
In order to reduce the readout noise, the measurements
were done using multiple correlated readout mode, with 4 readouts
before and after the integration. We used a slit width of $0\farcs75$ or 
1\arcsec, depending on the seeing conditions, which yielded
a spectral resolution in the  range $R=$ 500--660 and 550--700 in 
the ZJ and HK spectra, respectively.
The wavelength calibration was provided by observations of Argon and Xenon 
lamps available in the calibration unit at the A\&G box of the telescope.
In order to obtain the telluric correction and the flux calibration,    
nearby A0V or G2V stars were observed with the same configuration as the object.

\begin{figure*}
\centerline{\includegraphics[angle=90,width=13cm]{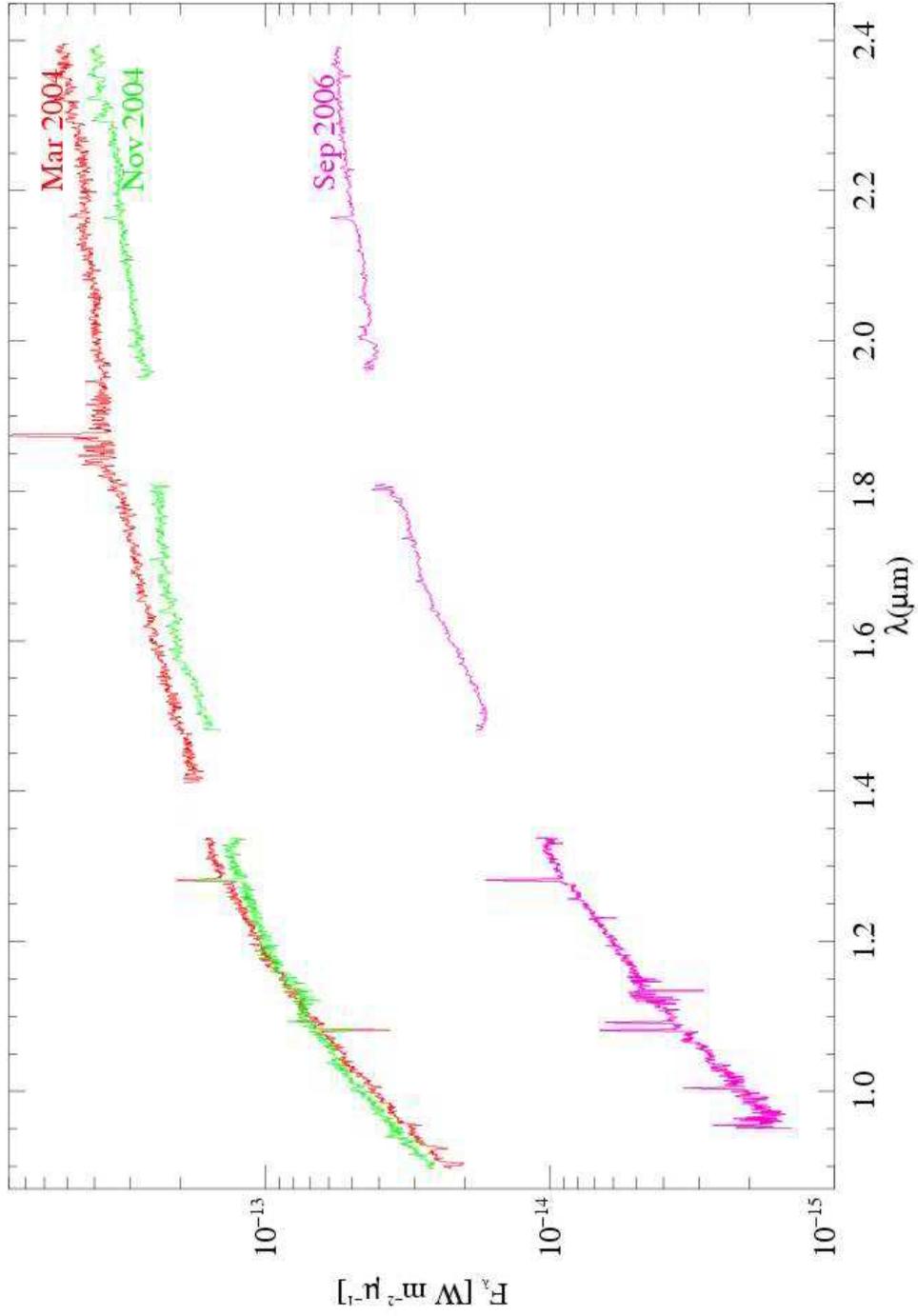}}
\caption{Flux calibrated near-infrared spectra of V1647\,Ori obtained 
with LIRIS on WHT.}
\label{Fig_sp_cal}
\end{figure*}

The data were reduced and calibrated using the package `liris$\_$ql'.  
Consecutive pairs of AB two-dimensional images 
were subtracted to remove the sky background, then the resulting images 
were wavelength calibrated and flat-fielded before registering and coadding 
all frames to provide the final combined spectrum. 
One dimensional spectra were extracted with the IRAF `apall' task.
The extracted spectra were divided by a composite
spectrum  to eliminate telluric contamination. 
This composite spectrum was generated from the 
observed spectra of the calibration star, divided by a stellar model and
convolved to our spectral resolution.
We used the publicly available code `Telluric' \citep{Vacca03} 
to perform the correction when A0 stars were observed. 
Differences in the strength of telluric features likely due to 
mismatch of air masses and variation of
atmospheric conditions between observations of the object and the reference 
star were taken into account using Beer's law.  The IRAF task `telluric'
was used in this step. The flux calibration was carried out normalizing
to the J, H, and K$_s$ magnitudes obtained in our near--IR photometry. 
We show in Fig.~\ref{Fig_sp_cal} the flux calibrated infrared spectra 
for those observing dates when the full range from 0.9 to 2.4\micron\ was
observed. 

\subsection{Optical spectroscopy}

Medium-resolution optical spectra were taken on three nights: 2004 October 28, 
November 1, and December 24, with the Double Beam Spectrograph mounted on
the 2.3\,m telescope of the Siding Spring Observatory (Australia). We used 1200
mm$^{-1}$ gratings in both arms of the spectrograph, while the exposure time
was set to 20 minutes. The slit width was 2\arcsec \ and the instrument's
rotator kept it in the parallactic angle. The image scale was 0\farcs9/pixel in
the direction of the  dispersion and 3\farcs6/pixel perpendicular to it. The
observed spectral ranges and resolutions were as follows:  blue -- 4800--5240
\AA, $\lambda/\Delta \lambda$=8300; red -- 5820--6780 \AA, $\lambda/\Delta
\lambda$=6500. All spectra were reduced with standard IRAF tasks, including
bias and flat field corrections, aperture extraction and wavelength
calibration. 

In the blue spectral range we observed a barely visible continuum, and 
could not detect the absorption spectrum observed by \citet{Briceno04}
in  2004 February. These blue spectra were excluded from the
further analysis. In the red range we observed the H$\alpha$ 
emission line on a featureless continuum.  
The equivalent widths of the stellar H$\alpha$ line determined from
our observations are 32, 34 and 27\,\AA\ for the dates 2004 Oct 28, 
Nov 1 and  Dec 24, respectively. 

\section{Brightness variations during the V1647\,Ori outburst}
\label{sc:outburst_hist}

\subsection{Long term evolution}

\begin{figure*}
\centerline{\includegraphics[width=18cm]{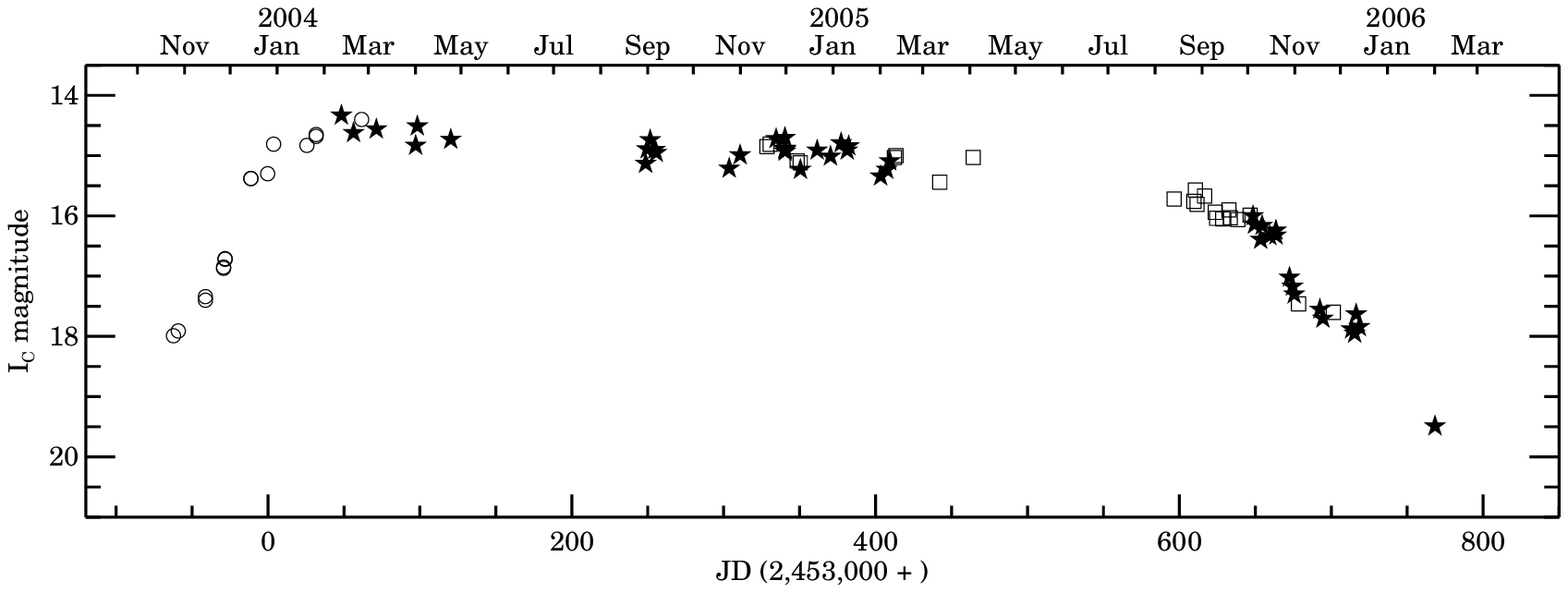}}
\vskip 3mm
\caption{{\it I\/}$_\mathrm{C}$-band light curve of V1647\,Ori
    covering the complete 2-years outburst period. {\it Filled
    stars}: our observations reported in Tab.~\ref{tab:opdata}; {\it open
    circles}: data from \citet{Briceno04}; {\it open squares}: data from \citet{Ojha06}. 
    The uncertainties are comparable with symbol sizes.}
\label{fig:ilight}
\end{figure*}

In order to examine the flux evolution during the
whole outburst history we combined our {\it I}$_\mathrm{C}$ measurements
with those of \citet{Briceno04}, which covered the beginning of
the eruption. The complete {\it I}$_\mathrm{C}$-band light curve is
displayed in Fig.~\ref{fig:ilight}. Following an initial brightening of
$\sim 4.5\,$mag between 2003 November and  2004 February \citep{Briceno04},
the star faded only moderately during the next one and a half
years. This gradual decrease of brightness was observable in all photometric
bands with similar rates (Fig.~\ref{fi:allbphot}). Then, between
2005 October and  November the brightness of the star suddenly dropped by
more than 1 mag at all optical wavelengths \citep{Kospal05}. This decay 
was also monitored  by \citet{Semkov06} and \citet{Ojha06}.
The rapid fading continued in the next months, although the 
rate was slightly slower after 2005 mid--November.  By 2006 February, some 800
days after the onset of the eruption, V1647\,Ori was very close to
the reported pre-outburst state.

The complete light curve of the outburst can naturally be divided into three 
phases: the initial rise, the ``plateau'', and the final decay. In order to
quantify the rate of brightness evolution, we fitted straight lines to the 
data points in Fig.~\ref{fig:ilight} for each phase separately. The results
are $-1.3$~mag/month, 0.04 mag/month and 0.8~mag/month, respectively. It is
remarkable that the pace of brightness change in the rising and decaying
phases is similar. 

The fact that the fading rate of V1647\,Ori is apparently wavelength 
independent is different from the case of another eruptive star
V1057 Cyg, whose decline rate decreased monotonically from the ultraviolet
to the mid--infrared during the years following the outburst 
\citep{KH91}. However the 
decline of V1057 Cyg was most pronounced in the B band, which was not observable
in the case of V1647\,Ori.

\subsection{A periodicity of 56 days}

The light curves in Figs.~\ref{fi:allbphot} and \ref{fig:ilight} show that, 
in addition to the slow decline during the high plateau, 
short time scale variations can be seen in each
photometric band.  Definite brightness minima can be recognized in
each optical band e.~g. on 2004  December 10 (JD 2,453,350) and on 2005 February 1  
(JD 2,453,403).  Similar depressions can be seen  around 
2004  March 7 (JD 2,453,072) and 2004  April 12 (JD 2,453,108) 
in the light curves presented by \citet{Walter04}.
Visual inspection suggests a possible periodicity in the light curve.
In order to study this phenomenon quantitatively,  
a period search analysis was conducted, using the programme package {\sl
MUFRAN\/} developed by \citet{Kollath}. 
The {\sl MUFRAN\/} software is based on the Fourier transform.
For a recent review of the time-series
analysis of variable star data, see \citet{Templeton} and references therein.

\begin{figure*}
\centerline{\includegraphics[angle=90,width=5cm]{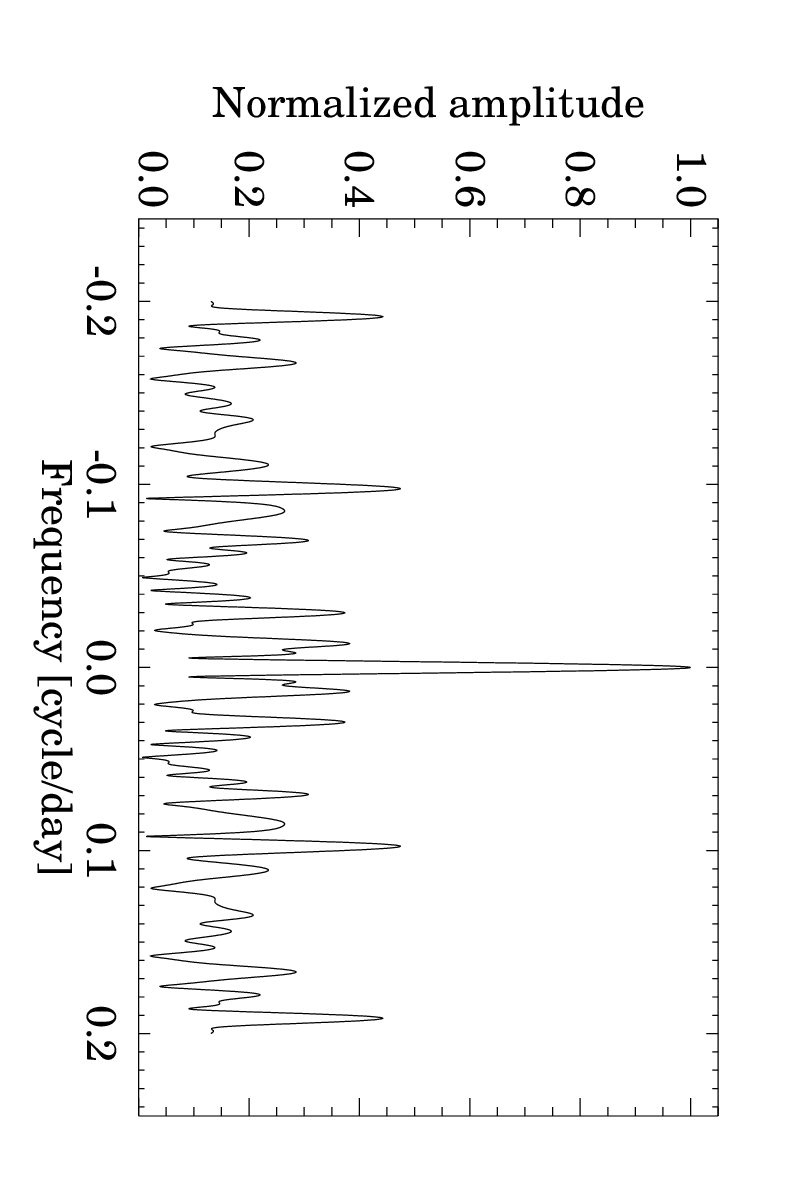}\hspace{5mm}
\includegraphics[angle=90,width=5cm]{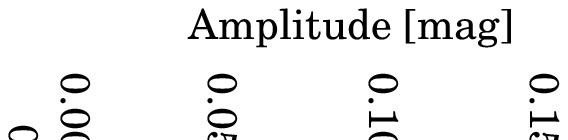}\hspace{5mm}
\includegraphics[angle=90,width=5cm]{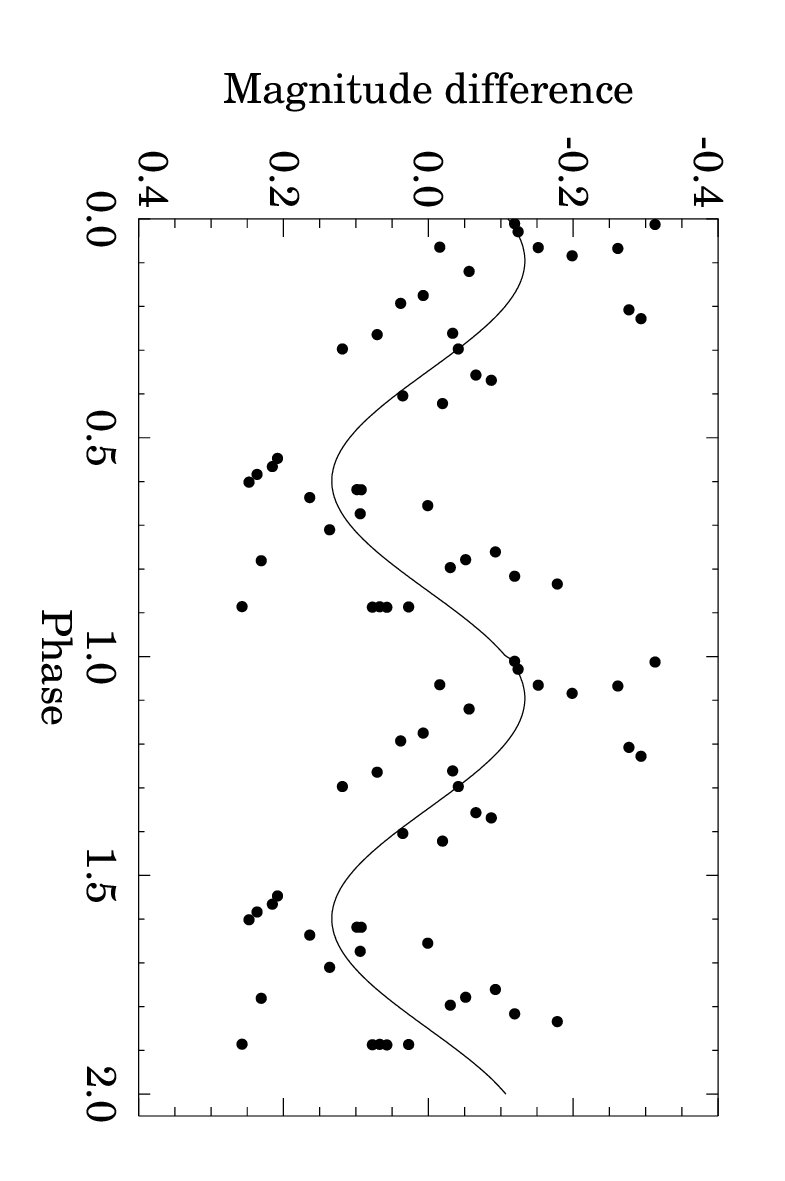}}
\caption{Results of the period search. {\sl Left\/}: The spectral
window function for the {\it I\/}$_\mathrm{C}$ photometric sample;
{\sl Middle\/}: A part of the Fourier power spectrum of the $I_C$
data. The first high peak at the frequency of 0.0179~c/d corresponds
to a period of 55.87~days.  {\sl Right\/}: The $I_C$ residual phase curve
folded on the period of 55.87~days.}
\label{Fig_mufran}
\end{figure*}

\begin{figure}
\centerline{\includegraphics[width=7cm]{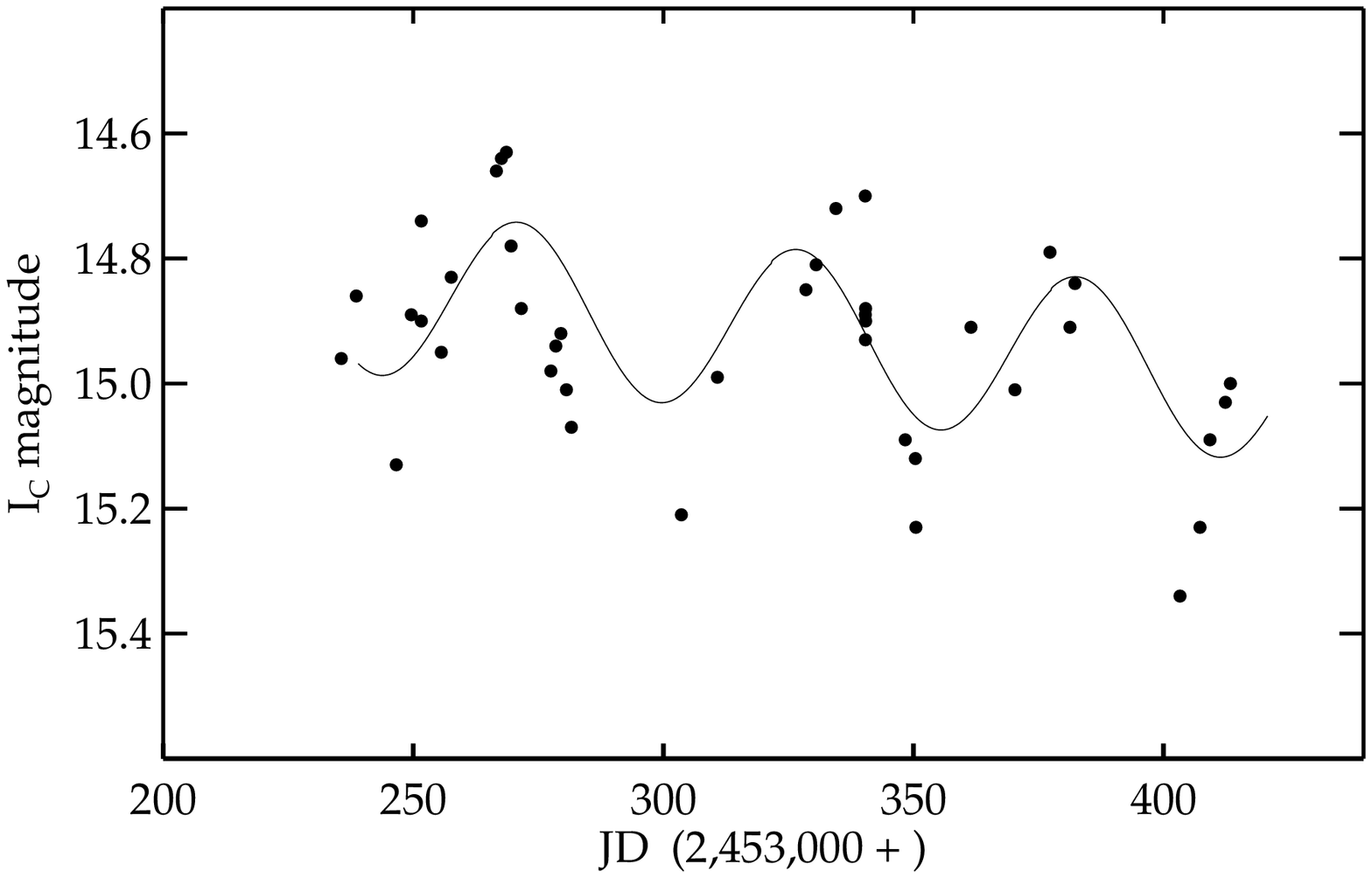}}
\caption{The $I_C$ light curve with the fitted sinusoidal curve overimposed.
The dimming trend was also included. The quasi--periodic variation of the 
light curve is clearly detected.}
\label{Fig_fitsinus}
\end{figure}

Our $R_C$ and $I_C$ data listed in Tab.~\ref{tab:opdata} were supplemented with
data taken from \citet{Semkov04a, Semkov06} in the $R_C$ and $I_C$ filters.
The imported data were zero offset by 0.42 ($R_C$) and 0.17 magnitudes ($I_C$), 
in order to reach a common average magnitude scale with our data.
The most likely reason of such difference is the use of different methods 
to extract the stellar flux.  
Only the time interval between 2004 Aug 18 and 2005 February 11 
(JD\,2,453,235--JD\,2,453,413) was considered because outside
this interval the longer time-scale variation may falsify the result
of the period analysis.
Due to the evidence of a long term brightness variation,
allowance was made for the monotonous dimming: a straight line was fitted to
each dataset and subtracted to form magnitude differences with respect to this
line.   
The period search was performed separately for data in each
photometric band. In the case of finite dataset and unevenly sampled
data some false frequencies (aliasing) can appear in the Fourier power
spectrum. These alias frequencies are centered on the real signals
offset from those peaks as indicated by the spectral window function
of the Fourier transform.  The temporal distribution of the
photometric data of V1647~Ori is favourable and the spectral window 
(see the left panel of Fig.~\ref{Fig_mufran}) indicates a negligible aliasing
caused by the data sampling. Here we only show the spectral window
function corresponding to the $I_C$ filter dataset, and note that the shape
and the alias structure is practically the same in the case of the
spectral window for the $R_C$ dataset.
The power spectra typically contain a number of peaks (see middle panel 
of Fig. \ref{Fig_mufran}).  
We show in this figure only the lowest frequency part of the power spectrum, 
the rest of the power spectrum only contains much lower peaks than seen here. 
The highest peak appears at the 
frequency of $f=0.0179$ cycle/day (equivalent to a period
of 55.87 days) in the {\it I\/}$_\mathrm{C}$ data.  
The S/N ratio is slightly above 3 in the region of
the $f = 0.0179$ c/d peak, which gives significance
to this peak. The S/N value was determined as the ratio of
the amplitude of the main peak and the average amplitude present in
the power spectrum after prewhitening with this frequency. 
Moreover, a peak can be considered as physically meaningful, 
if it appears in the power spectra of each photometric observational series. 
The corresponding peak is situated at the frequency of 0.0163 c/d 
in the analysis of  the shorter series of the {\it R\/}$_\mathrm{C}$ 
filter data.
The phase curve of the {\it I\/}$_\mathrm{C}$ residual photometric data 
folded on the 55.87 day period 
is shown in the right panel of Fig.~\ref{Fig_mufran}. 
The {\it R\/}$_\mathrm{C}$ band folded
light curve is qualitatively similar to that shown in
Fig.~\ref{Fig_mufran}. It is worth mentioning that the negligible phase
difference between the {\it R\/}$_\mathrm{C}$ and {\it
I\/}$_\mathrm{C}$ folded light curves also implies a common origin of
this periodicity. This means that the $\sim 56$ day periodicity is
confirmed. We have plotted in Fig.~\ref{Fig_fitsinus} the portion of
the  $I_C$ light curve which was fitted with the sinusoidal 
curve plus a monotonic decrease. The time interval covered by 
our data is $200$~days, which is only about 4 cycles of the
possible period of 56 days. The periodic behaviour is clearly 
seen in the curve, although a sinusoidal variation is regarded only as a 
first and crude approximation.

\subsection{Hour scale variations}

\citet{Ojha05} claimed weekly variations 
of the optical flux of V1647~Ori.
\citet{Grosso} reported on a periodic variation in the X--ray flux 
of 0.72 days, about 17 hours.  
We have also searched for very short term variability in our
optical data.  
We compared 5 consecutive {\it I\/}$_\mathrm{C}$-band observations
of V1647\,Ori obtained on the same night (2004 November 30) during a
period of 2.5 hours with the 1~m RCC telescope. The first four 
measurements are identical within their formal photometric
uncertainties ($\approx\,0.03\,$ mag), but the fifth measurement 
deviates about 0.2\,mag from the others. This figure 
may be the characteristic variation expected on hour scale.

\subsection{The possible origin of variability}

\begin{figure*}
\centerline{\includegraphics[width=8cm]{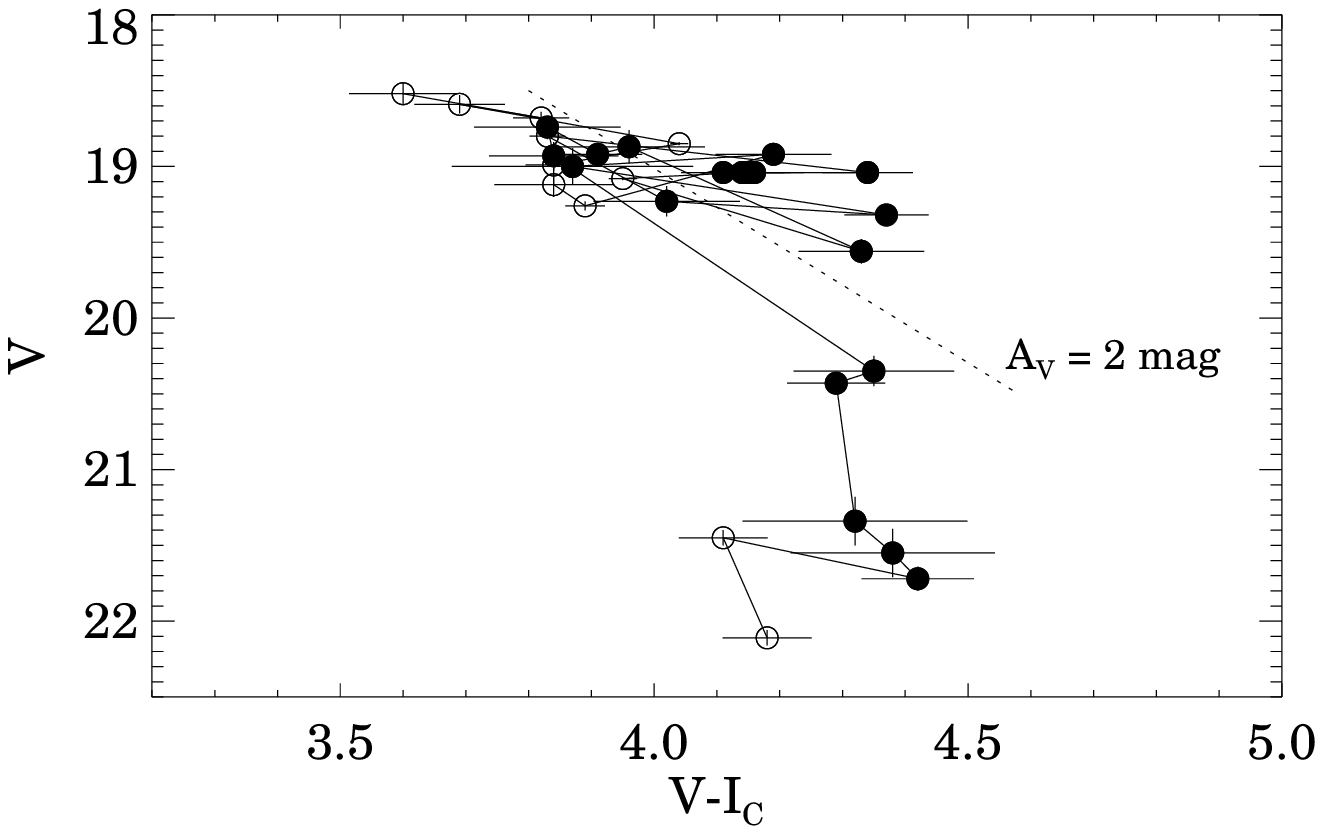}\hspace{5mm}
\includegraphics[width=8cm]{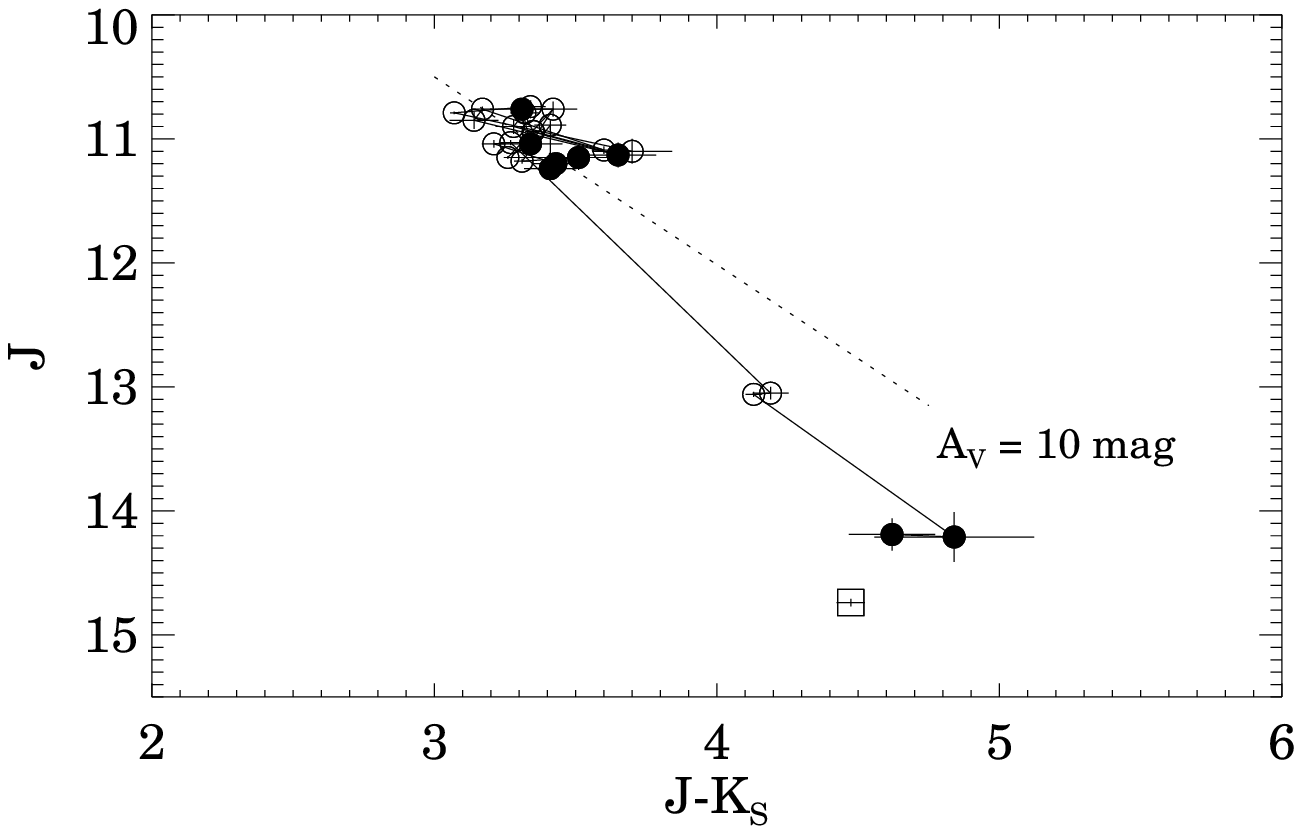}}
\caption{Relationship between the color indices and magnitudes of
V1647~Ori during the monitoring period.  
{\sl Left\/}: {\it V\/} {\it vs.}  $V -I_\mathrm{C}$; 
{\sl Right\/}: {\it J\/} {\it vs.} $J - K_\mathrm{s}$. The filled 
symbols represent our measurements, and the open symbols are taken
from \citet{Ojha06}. The dotted
lines indicate the slope of the normal interstellar reddening \citep{CFPE}.
Solid lines connect consecutive observed points.}
\label{fi:col_mag}
\end{figure*}

In order to study further the mechanisms causing the initial brightening 
and the final fading, we plotted in Fig.~\ref{fi:col_mag}
optical and near-IR color--magnitude diagrams, and looked for
relationships between brightness and color index variations. 
We also marked 
in Fig.~\ref{fi:col_mag} the standard interstellar reddening law \citep{CFPE}.  

\citet{RA} noted that the star moved along the extinction path in the J--H vs.
H--K diagram during the initial brightening phase. 
\citet{Ojha06} demonstrated that the final 
fading of the object in November 2005 also followed the same path in 
the opposite direction. 
However the variation of the near-infrared and visible colors
is not compatible with simple dust obscuration. It can be seen from 
Fig.~\ref{fi:col_mag} that the amount of extinction needed to explain the
color variation in the near-infrared is much larger than the one required
in the visible. In order to explain the amplitude of the outburst a color
independent flux increase has also to be assumed. 
A similar conclusion was reached by \citet{McGehee}. 

In order to explain the flux variations during the plateau phase  we looked at
any possible relationship between brightness and  color index variations 
(Fig.~\ref{fi:col_mag}).  A loose trend that the star tends to be redder when
dimmer  can be observed in the diagrams. \citet{Eiroa} successfully explained 
the light variations of several young stars  by obscuration by dense slabs within
the circumstellar  disk.  Thus obscuration by circumstellar dust structures
orbiting at a distance of $R = 0.28 (M / M_{\sun})^{1 / 3} AU$, implied by our
$\sim 56$ day period,  may explain at least part of the photometric variations
during the plateau phase.

\subsection{Inclination of the star--nebula axis derived from the brightness evolution}
\label{Sect_incl}

\begin{figure}
\resizebox{\hsize}{!}{\includegraphics{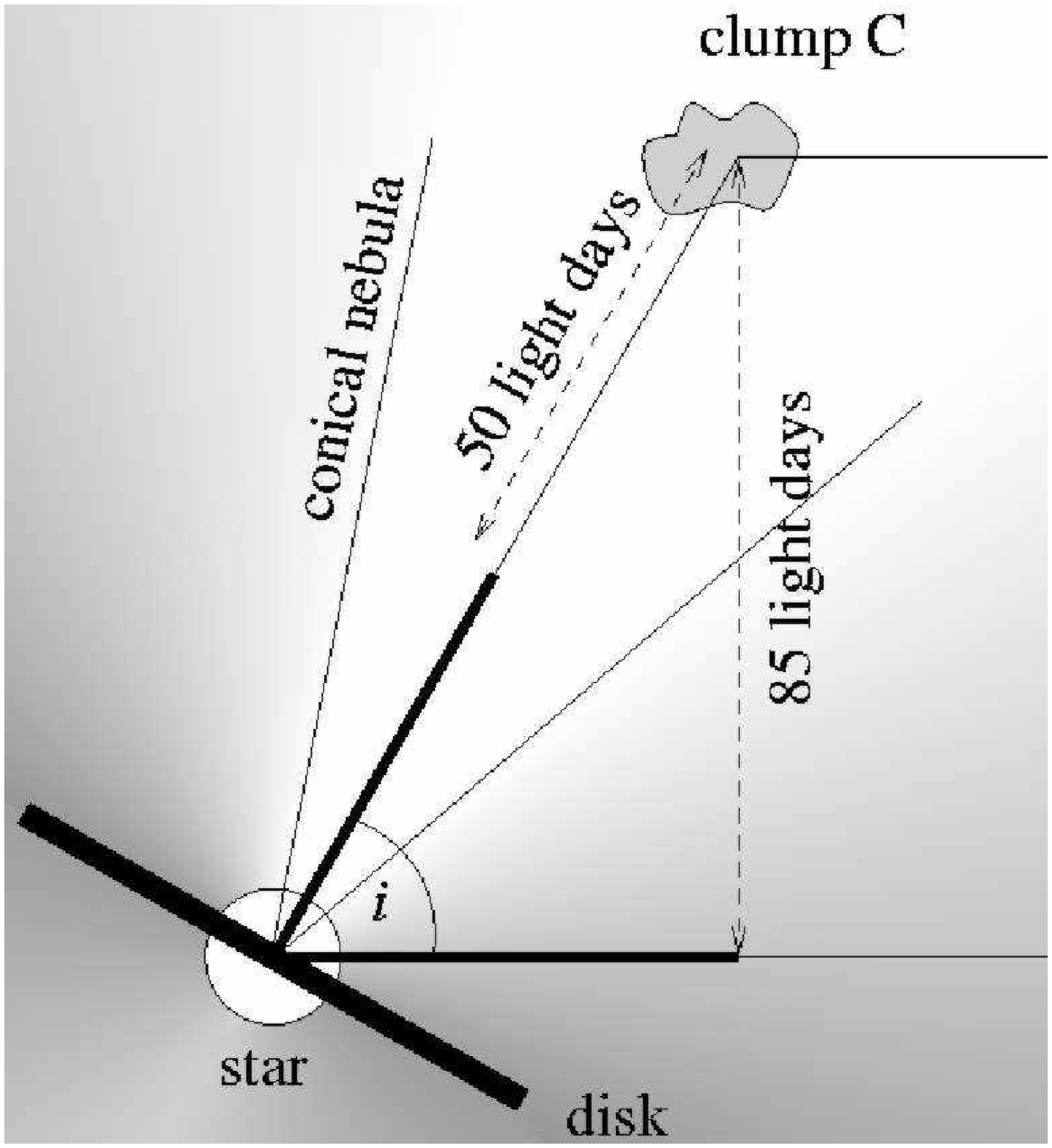}}
\caption{Geometry of McNeil's Nebula, deduced from the time delay between 
the light curves of V1647~Ori and Clump~{\it C\/}.}
\label{Fig_incl}
\end{figure}

McNeil's Nebula shines due to the scattered light of V1647~Ori, therefore its 
observed properties are closely related to the brightness of the star. 
\citet{Briceno04} examined the temporal evolution of 
two well-defined positions of the nebula (marked as {\it B\/} and {\it C\/}
in their fig.\,2), 
and concluded that their light curves followed the 
brightening of the star with delays of about 14 days (Pos.~{\it B\/})
and of more than 50 days (Pos.~{\it C\/}). The angular distance of 
36\arcsec\ between the
star and Pos.~{\it C\/} implies a delay of 85 days. Since these delays 
carry information on the inclination of the nebula (see Fig.\,\ref{Fig_incl}),
we reanalysed the three {\it I\/}$_\mathrm{C}$-band light curves (from 
Fig.~3 in \citealp{Briceno04}), supplemented 
with our measurements obtained in the period 2004 February--April, in order 
to find the accurate time shifts between them.  
First we interpolated the light curves for
a time sampling of 0.25~days. The resulting light curves 
of V1647\,Ori, Pos.~{\it B\/}, and Pos.~{\it C\/} are
called $A(t_i)$, $B(t_i)$ and $C(t_i)$, respectively, where $t_i$ is the time
($i=1...N$). Then we calculated a function called $D(s)$ as follows:
$$ D(s) = \sum_{i=1}^N A(t_i) - C(t_i-s) $$
The  quantity $s$ which results in the lowest value of $D(s)$ gives the time shift
between the light curves of {\it A\/} and {\it C\/}. The shifts
between {\it A\/} and {\it B\/} can be calculated similarly. 
In this manner we obtained a time shift 12.5$\pm2$ days between V1647\,Ori and 
{\it B\/} and 50$\pm3$ days between V1647\,Ori and Pos.~{\it C\/}. 

The measured time delay between the light curves of V1647\,Ori and Pos.~{\it C\/}, 
together with the projected separation of the two objects results in an angle of
61$^{+3}_{-2}$ degrees between the  axis of the nebula and
the line of sight. Because the clump lies inside the polar stellar wind cavity
of the star, this angle is a good estimate for the inclination 
of the rotation axis of the star (see Fig.~\ref{Fig_incl}). 
We estimated the uncertainty
of the inclination derived this way as follows. We assumed that the Herbig--Haro object
{\sl HH\,23\/} \citep{EM} lies exactly 
along the rotation axis of the star. Then we assumed that the projected 
angle between the {\sl star--Pos.~~C\/} and {\sl star--HH\,23\/} directions,
11\degr, is the same as 
the difference in their tilts to the line of sight. Thus the total uncertainty of 
the inclination derived from the time delay of the brightness variation of clump 
{\it C\/} is $\pm14\degr$.

The time delay between the light curve features of V1647\,Ori 
and Pos.~{\it B\/} results in 65\degr $\pm8\degr$. The inclinations derived
from positions {\it B\/} and {\it C\/} are compatible within the errors. 
Due to the larger opening angle of the conical nebula at small 
distances from the star, position {\it B\/}
is less suitable for estimating the inclination of the stellar rotation axis.

This result apparently conflicts with \citet{Rettig} finding
that the disk of V1647~Ori is seen nearly face-on, because the absorption line
of the cold CO due to the outer, flared regions of the disk cannot be observed 
in the high resolution infrared spectrum of the star.
But if the thickness $H$ of the disk at a distance $R$ from the 
star does not exceed $H / R  \approx 0.50$, then  the absorption lines produced 
by the outer disk is not expected to be observed in the spectrum of the star.

\section{Spectral evolution}
\label{sec:specevol}

\begin{figure*}
\centerline{
\includegraphics[width=10.0cm]{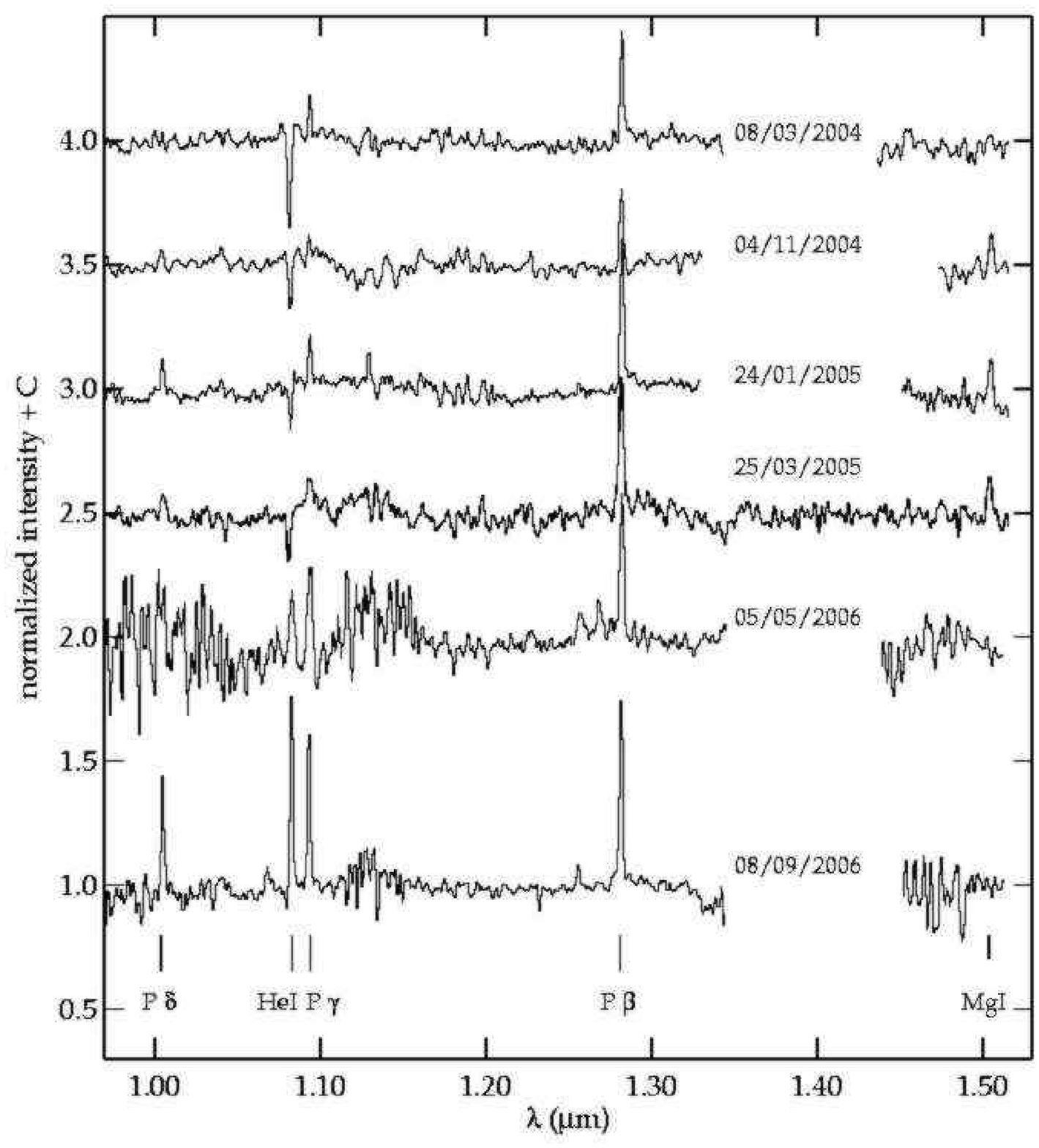}}
\vskip 8mm
\caption{\label{Fig_sp_zj} Near-infrared  ({\it ZJ\/} band) spectra of V1647 Ori obtained 
with LIRIS, normalized to the continuum.} 
\end{figure*}

\begin{figure*}
\centerline{
\includegraphics[width=10.0cm]{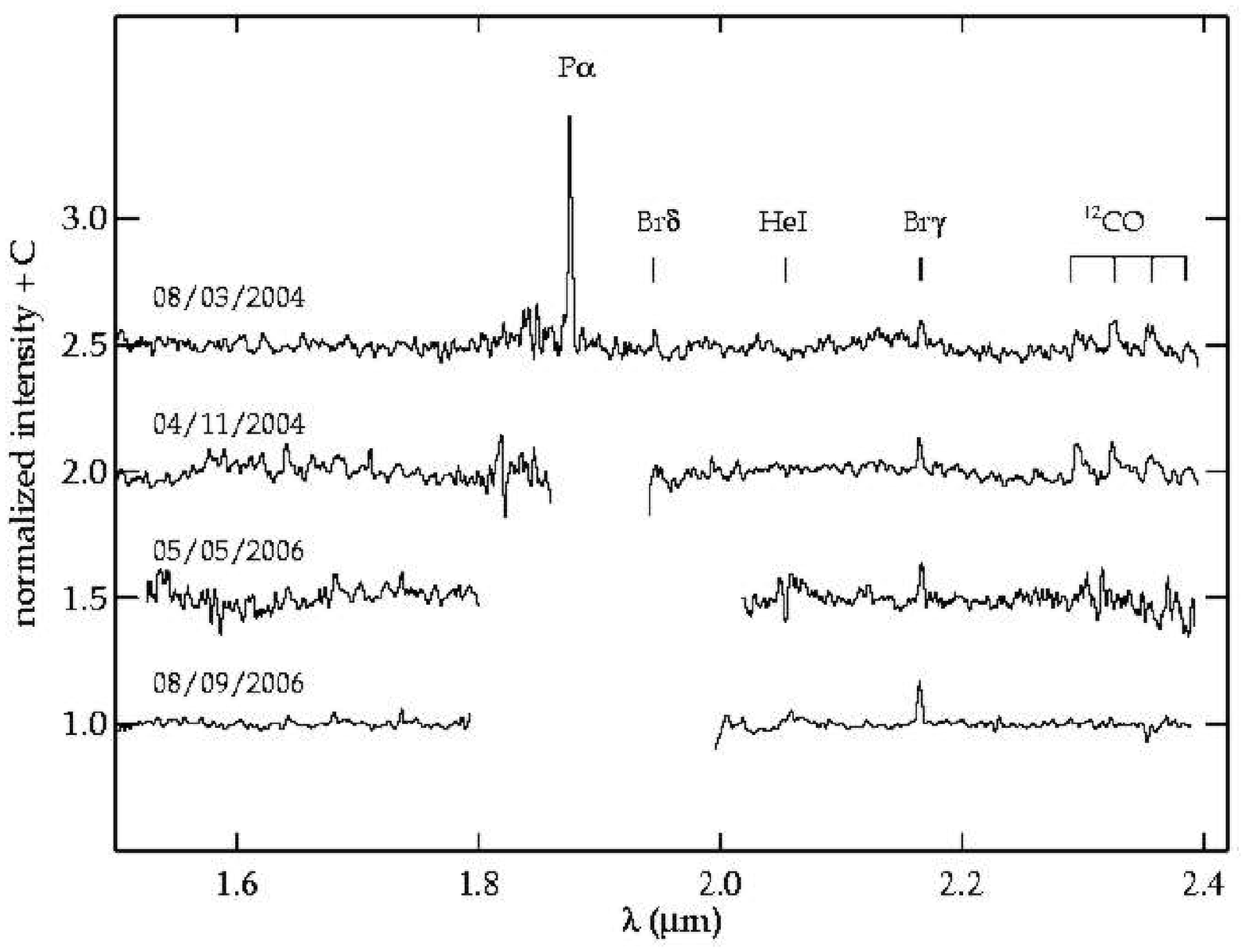}}
\vskip 8mm
\caption{\label{Fig_sp_hk} Near-infrared  ({\it HK\/} band) spectra of V1647 Ori obtained 
with LIRIS, normalized to the continuum.} 
\end{figure*}

Our near-infrared spectra, normalized to the continuum, are displayed in 
Fig.~\ref{Fig_sp_zj} (ZJ band) and Fig.~\ref{Fig_sp_hk} (HK band).  The most
prominent features are the Pa$\beta$, Pa$\gamma$, and Br$\gamma$ lines of HI 
in emission, the HeI 1.083\,\micron\  line most of the time seen in absorption,
the  MgI 1.50\,\micron\ line, and the rovibrational $\Delta \nu = +2$ transitions  of CO at
2.3--2.4\,$\mu$m in emission. 
These spectral features are present in all spectra,  though significant
changes can be recognised in the line strengths during our observational period
of 30 months. In the following we analyse these spectral variations, including
also  data sets from \citet{Walter04}, \citet{Gibb06} and \citet{Ojha06}, which
cover  shorter periods.  We note that our last spectra were taken in 2006 May
and Sep when the outburst was over, thus probably they are representative 
of the quiescent phase of the  object. 

\begin{figure}
\centerline{
\includegraphics[width=9truecm]{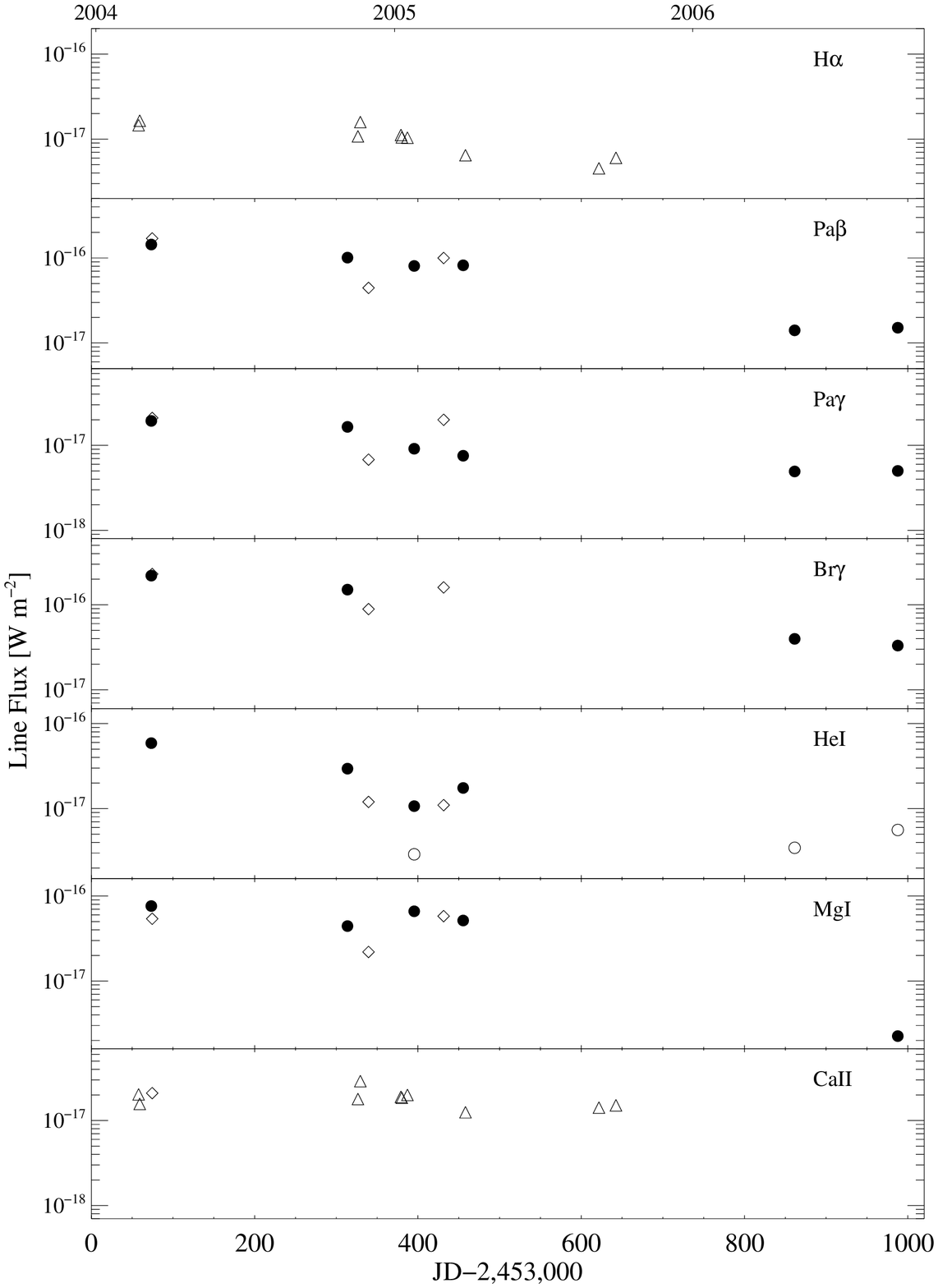}}
\vskip 7mm
\caption{\label{Fig_flux} Flux evolution of near-infrared and optical 
spectral lines of V1647\,Ori. {\it Circles:} this study, open circles 
correspond to HeI emission component  
(Tab.\,\ref{Tab_nir_lines}); {\it diamonds:} data
taken from \citet{Gibb06}; {\it triangles:} data taken from 
\citet{Walter04}  and \citet{Ojha06}. The formal error bars 
are smaller than the symbol sizes.} 
\end{figure}

Table~\ref{Tab_nir_lines} lists the equivalent widths and fluxes of
the infrared lines. The near-infrared line fluxes were computed
from the absolutely calibrated spectra, with a  
typical estimated uncertainty of about 15\% in the brightest lines to 
25\% in the faintest ones.
Figure~\ref{Fig_flux} shows the time evolution of the line fluxes. 
In addition, we plotted the H$\alpha$ and CaII\,$\lambda$\,8542\,\AA{} 
lines, where line fluxes were derived by combining the equivalent 
widths published by \citet{Walter04} and \citet{Ojha06},
and $I_C$ and $R_C$ magnitudes taken from Tab.\,\ref{tab:opdata} 
or from the literature for the same day as the spectroscopic data. 
The typical uncertainty of these latter flux values, derived from the 
errors of the magnitudes and equivalent widths, is about 6~percent.
We noticed inconsistencies when 
comparing our line fluxes with those published by other authors for 
close observing dates. For this reason we recomputed 
their line fluxes combining the reported EW with  photometric data. These 
are the values plotted in Fig.~\ref{Fig_flux}.
 

\begin{deluxetable}{lcrrrrrrrrrrrr}
\tablewidth{0pt}
\tablecolumns{9}
\tablecaption{\label{Tab_nir_lines} Equivalent width and flux of near-infrared lines.
}
\tablehead{
\colhead{Line} & \colhead{$\lambda$ ($\mu$m)} &  \multicolumn{2}{c}{2004 Mar} & 
\multicolumn{2}{c}{2004 Nov} & \multicolumn{2}{c}{2005 Jan} &  
\multicolumn{2}{c}{2005 Mar} & \multicolumn{2}{c}{2006 May}  
& \multicolumn{2}{c}{2006 Sep}\\
\colhead{} & \colhead{} & \colhead{EW\tablenotemark{a}} & \colhead{Flux\tablenotemark{b}} &  \colhead{EW} & \colhead{Flux}
& \colhead{EW} & \colhead{Flux} & \colhead{EW} & \colhead{Flux} & \colhead{EW} & \colhead{Flux} & \colhead{EW} & \colhead{Flux} }
\startdata
Pa$\delta$ & 1.005 & $-0.8$   & 1.2    &  $-2.5$   & 1.2     &  $-2.1$   & 0.5     &  $-1.7$   & 0.4     & \nodata & \nodata & $-12.4$   & 0.26    \\
HeI        & 1.083 &  9.18  & $-5.9$   &   4.3   & $-3.0$    &  3.3    & $-1.1$    &   5.2   & $-1.8$    & $-11.6$   & 0.3     & $-17.0$   & 0.56    \\
Pa$\gamma$ & 1.094 & $-3.0$   & 1.9    &  $-2.4$   & 1.7     &  $-2.7$   & 0.9     &  $-2.2$   & 0.8     & $-14.8$   & 0.5     & $-13.4$   & 0.50    \\
Pa$\beta$  & 1.282 & $-10.7$  & 14.4   &  $-8.5$   & 10.1    &  $-12.5$  & 8.1     & $-15.1$   & 8.2     & $-14.9$   & 1.4     & $-16.7$   & 1.51    \\
MgI        & 1.504 & $-3.7$   & 7.6    &  $-2.8$   & 4.4     &  $-5.9$   & 6.6     &  $-6.7$   & 5.2     & \nodata & \nodata & $-1.3$    & 0.23    \\
Br$\delta$ & 1.945 & $-2.9$   & 10.5   & \nodata & \nodata & \nodata & \nodata & \nodata & \nodata & \nodata & \nodata & \nodata & \nodata \\
HeI        & 2.058 & $<0.8$ & $<3.0$ &  0.8    & $-2.5$    & \nodata & \nodata & \nodata & \nodata & \nodata & \nodata & $-1.7$    & 0.77     \\
Br$\gamma$ & 2.166 &  $-5.2$  & 22.0   & $-4.8$    & 15.1    & \nodata & \nodata & \nodata & \nodata & $-6.9$    & 4.0     & $-6.7$    & 3.31     \\
\enddata
\tablenotetext{a}{Units \AA}
\tablenotetext{b}{Units $10^{-17}\mathrm{Wm}^{-2}$}
\end{deluxetable}

\paragraph{Hydrogen lines.} Figure~\ref{Fig_flux} shows that the line fluxes of
Pa$\beta$, Br$\gamma$  and H$\alpha$  declined slowly during the period
of observations. The Pa$\gamma$ shows a similar trend, but with higher scatter.

In many young stellar objects the origin of HI lines is related to accretion
processes. \citet{Muzerolle98a} established empirical relationships between 
the accretion rate and the luminosity of Pa$\beta$ and Br$\gamma$ lines.  
The data in Fig.~\ref{Fig_flux} clearly indicate parallel  decrease of the 
strongest hydrogen emission lines during the whole outburst, which would 
suggest a decreasing accretion rate. 
Application of the relationships of \citet{Muzerolle98a} with the  assumptions
that V1647~Ori is located at 400~pc from the Sun; suffers an extinction of
$A_V=11$\,mag;  has a mass  of 0.5\,M$_{\sun}$ and an accretion radius of
3\,R$_{\sun}$,  results in an accretion rate about $5 \times 10^{-6}
M_{\sun}$/yr in 2004 March, and an order of magnitude lower value in 2006 May.

The Paschen lines of HI showed P\,Cygni profile with a blueshifted
absorption  only in 2004 March, whereas the absorption part disappeared in
later observations.
\citet{Gibb06} also reported the same variations in the profiles of 
the HI lines.  
Similar  changes of the H$\alpha$ and CaII lines were reported 
by \citet{Ojha06}. These results indicate
that at the beginning of the outburst the source of the HI   lines was at least
partly stellar wind, whose strength decreased  during the outburst. 
As a consequence, the drop of accretion rate was probably smaller than we
derived in the previous paragraph.

\citet{Walter04} and \citet{Ojha06} suggested a possible periodic behaviour in
the H$\alpha$ equivalent width with a period of roughly 50 days. Since this
period is similar to the one found in our $I_C$ and $R_C$ light curves (55.87
days, see Sect.~3.2), it is tempting to speculate about a  connection  between
the two cyclic phenomena. In order to check it, we folded all available 
H$\alpha$ equivalent width data with the period of the optical light curve, and  constructed a phase
curve. It was found that the EW(H$\alpha$) and $I_C$ values oscillate in opposite
phase, i.e. the equivalent width of the  H$\alpha$ line is the largest when the
central object is the dimmest. Since the equivalent width is the ratio of the 
true line flux to the continuum level, it is very likely that the periodic variations 
of the H$\alpha$ EW are not  intrinsic but due 
to the periodically changing continuum flux.

\paragraph{Helium lines.} The evolution of the HeI 1.083~\micron\ line tells 
an interesting story. It was seen in blueshifted absorption during the whole
outburst, although its strength decreased with time
\citep[Fig.~\ref{Fig_flux}, see  also ][]{Gibb06}. The flux decreases 
quickly towards the end of 2004, and later decreases again in 2006 May. 
In our spectra of 2006 May and Sep, taken in the 
quiescent phase this line was detected {\it in emission}, without any  apparent velocity 
shift (Fig.~\ref{Fig_sp_zj}). 
The transition of the HeI 1.083~\micron\  line  from
absorption to emission is a remarkable event, not seen in any other line in our
spectra. 
This line also shows a blueshifted ($\sim 500 \mathrm{km\ s}^{-1}$) absorption 
component in the spectrum taken in 2006 Sep. 

In young stellar objects the HeI 1.083\,\micron\ line is usually attributed 
to hot winds. The profile of
the HeI absorption line observed in V1647~Ori was similar to those seen in a
large sample  of strongly accreting T~Tauri stars \citep{Edwards06}, where the
whole velocity range of the HeI wind can be observed against the stellar 
continuum. In these cases  \citet{Edwards06} argue that the wind traced by the
HeI\,\,1.083\,$\mu$m line originates close to the star and not from the inner
part of the circumstellar disk. This may also be the case for the V1647\,Ori 
system. The hard X-ray emission from V1647\,Ori, observed
by the X-ray observatories  {\sl Chandra\/} and {\sl XMM-Newton\/}
\citep{Kastner, Grosso}, indicates the presence of a magnetic reconnection ring
located at 1--1.5 stellar radii above the stellar surface in the plane of the
disk. We speculate on that the source of the wind is the
magnetic reconnection ring. 
This hypothesis is supported by the fact that
the strongest HeI line, observed on 2004  March~8,
\citep[Fig.~\ref{Fig_flux},][]{VCS, Gibb06}, was preceded by a high X-ray 
emission event on 2004 March~7, \citep{Kastner}, indicating a direct connection
between the two phenomena.
The  drop of the flux of the HeI 1.083\,\micron\
line  during the outburst suggests weakening of the wind. 

\citet{Edwards06} proposed that there may be two types of the hot wind,
producing absorption or emission of the HeI\,1.083\,$\mu$m line.
The transition of this line from absorption to emission  
at the end of the outburst indicates that the nature of the wind 
in V1647\,Ori changed between the two types of hot wind.

We also detected the HeI\,2.058\,$\mu$m line  in absorption both in 2004 November
and 2006 May. The apparent increase of its strength in the latter
spectrum (taken during quiescence) is due to the lower continuum. 

\paragraph{Other spectral features.}

In Fig.~\ref{Fig_flux} we also plotted the flux of the 
MgI\,1.505\,$\mu$m  emission line. The feature was detectable during the whole
outburst, showing hints for a slowly decreasing trend in flux. In the
spectrum of May 2006, during the quiescent phase, the line was not visible; the derived upper limit
in Tab.\,\ref{Tab_nir_lines} indicates that the 
MgI\,1.505\,$\mu$m emission is weak in the quiescent phase.

The last panel of Fig.~\ref{Fig_flux} shows the flux variation of the 
CaII\,8542\,\AA{} emission line based on literature data. 
This middle component of the CaII triplet was
found to be an excellent tracer of the accretion rate \citep{Muzerolle98b}. 
It can be seen that, contrary to all the other lines in Fig.~\ref{Fig_flux}, 
the flux of the CaII\,8542\,\AA{} line was  constant during the period of the
observations (there is no available measurement from the quiescent phase).  
A reason for the discrepancy may be that the hydrogen and calcium lines
originate from regions where the physical conditions are very different. 
\citet{Ojha06} noted that the density of the CaII region is very high 
and very optically thick. According to \citet{Muzerolle01} optically
thick lines are less reliable tracers of the variable accretion.
The accretion rate, derived from its luminosity, is about   $2 \times 10^{-6}
M_{\sun}$ / yr, close to the values obtained from the hydrogen lines for the 
high state of the outburst. 

The CO band head features at 2.3--2.4\,$\mu$m were observed in strong emission 
during the whole outburst period \citep[Fig.~\ref{Fig_sp_hk}, see also ][]{RA, VCS, Rettig, 
Gibb06}. However, the CO feature is not present in our quiescent spectrum 
suggesting that the appearance of this band is related to the outburst.

\section{The nebula around V1647 Ori}

McNeil's nebula is a reflection nebulosity which scatters the light of its
illuminating star, V1647~Ori.  The optical and near-infrared morphology  of the
nebula can be seen in Fig.\,\ref{fig:opt_examp} and Fig.\,\ref{fi:lirisJoverK},
respectively. Their study provides information  not only on the neighbouring
interstellar  medium but also on the outbursting star. In the following we
describe the morphology of the nebula  (see also \citealp{Ojha05, RA}).

Comparing the optical and near-infrared images one immediately notices that
the shape of the nebula looks rather  different at optical 
(Fig. \ref{fig:opt_examp}) and at infrared wavelengths  
(Fig.\,\ref{fi:lirisJoverK}).  In the optical images  the nebula shows a
conical shape, extending  up to a distance of $\sim 50\arcsec$ 
($\sim 22000$~AU at the distance of the nebula) to the north.  
The diffuse emission area
appears to be delimited by two bright regions indicating an opening angle of
$\sim 50^\circ$.  The north-west side is marked by region B 
\citep[named by][]{Briceno04}, a plume located at a distance between 5 and 10{\arcsec}.  
The north-east rim appears more extended and is delimited by region C 
\citep[also named by][]{Briceno04},  which is located around 35\arcsec\ from the star
and is elongated  perpendicularly to the radial axis from the star.  No
southern extension of the nebula can be recognized in the optical images.

In contrast to the optical, the infrared morphology indicates a much more 
compact  structure, though the cometary shape can still be seen with a sharp
curved  tail along the north--east side.  The nebula appears to be more 
extended in
the J band,  exhibiting diffuse emission both at the north--east and west sides. 
It appears roughly spherical in the K$_s$ band, although the structure is
flattened perpendicularly to the nebula axis; the largest size of
the structure is $\sim 20$\arcsec.  This size corresponds to a diameter of
$\sim 9000$~AU  at the distance of Orion B cloud, which is slightly smaller
than the size derived by  \citet{JFMM} for the matching submillimeter source
{\it Ori\,B\,N\,55\/}, about 13500~AU. The southward extension is also
clearly seen in the K$_s$ band.

\subsection{Color variations across the nebula}

\begin{figure*}
\centerline{
\includegraphics[width=9.0cm]{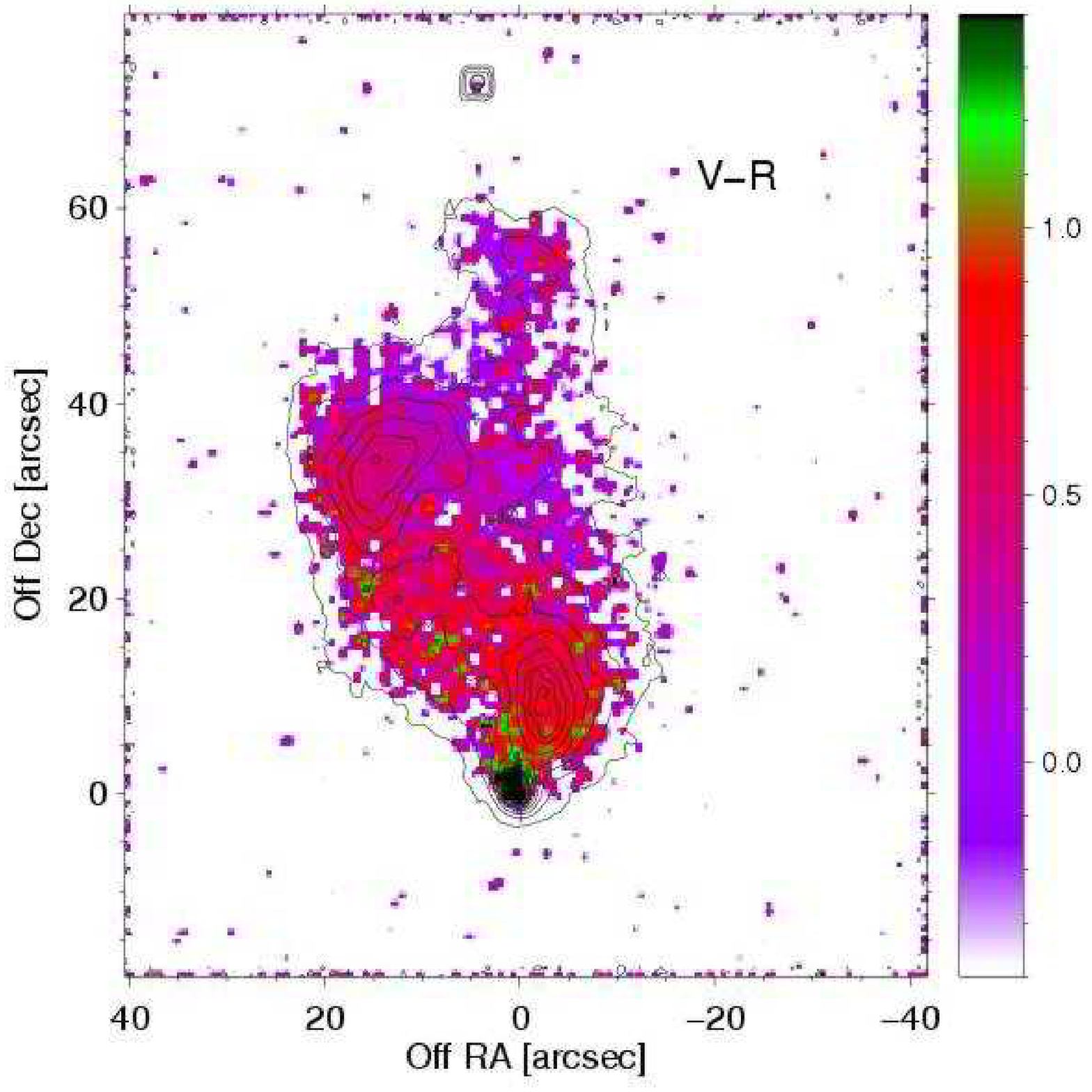}
\includegraphics[width=9.0cm]{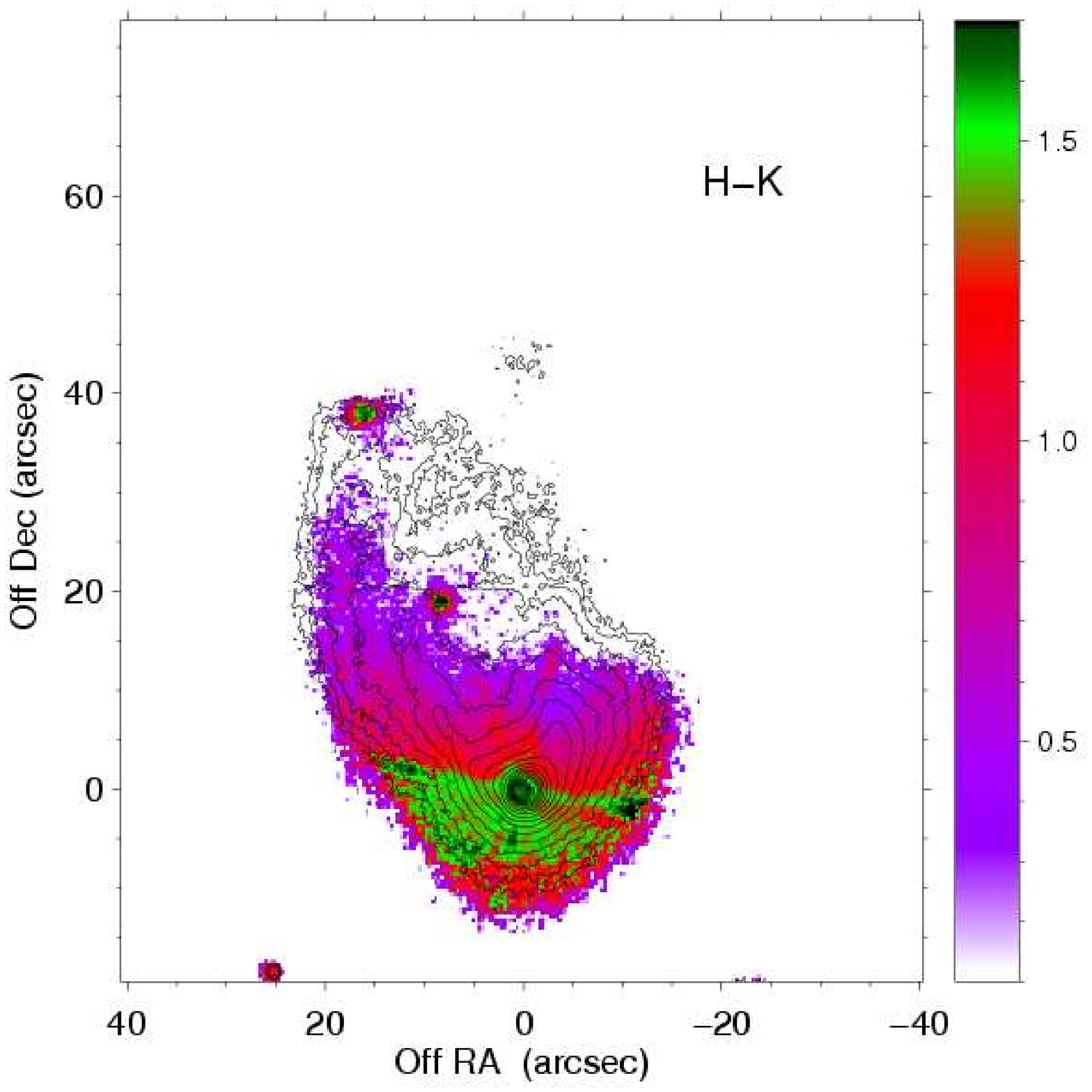}}
\vskip 3mm
\caption{\label{fi:cmaps} 
Color maps showing the $V-R_C$ ({\it left panel\/}) and $H-K_s$ ({\it right panel\/}) 
indices of the infrared nebula around V1647~Ori. The contours in the left and right panels 
represent the emission in $R_C$ and J bands, respectively.} 
\end{figure*}

\begin{figure*}
\centerline{
\includegraphics[width=7cm]{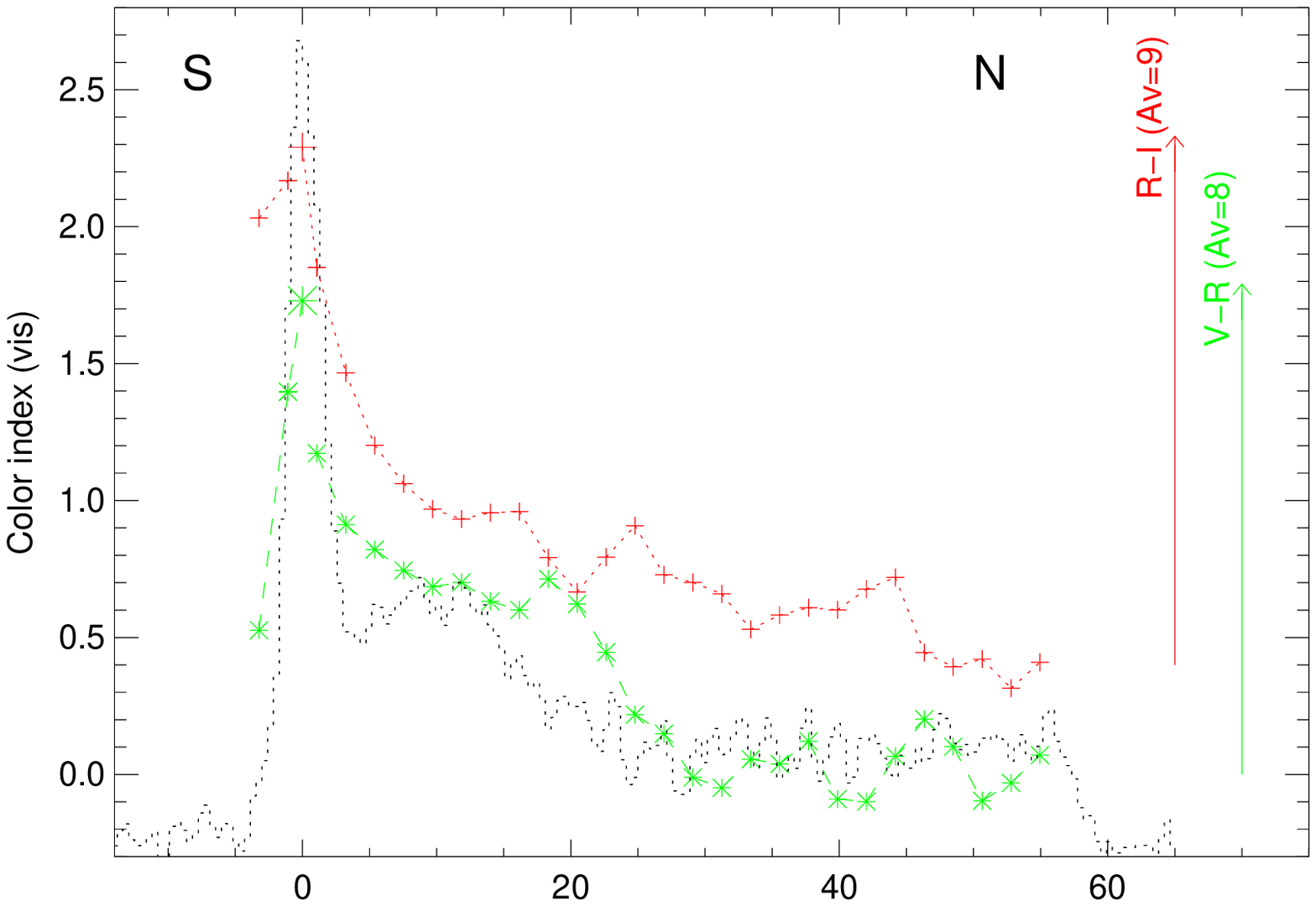}\hspace{10mm}
\includegraphics[width=7cm]{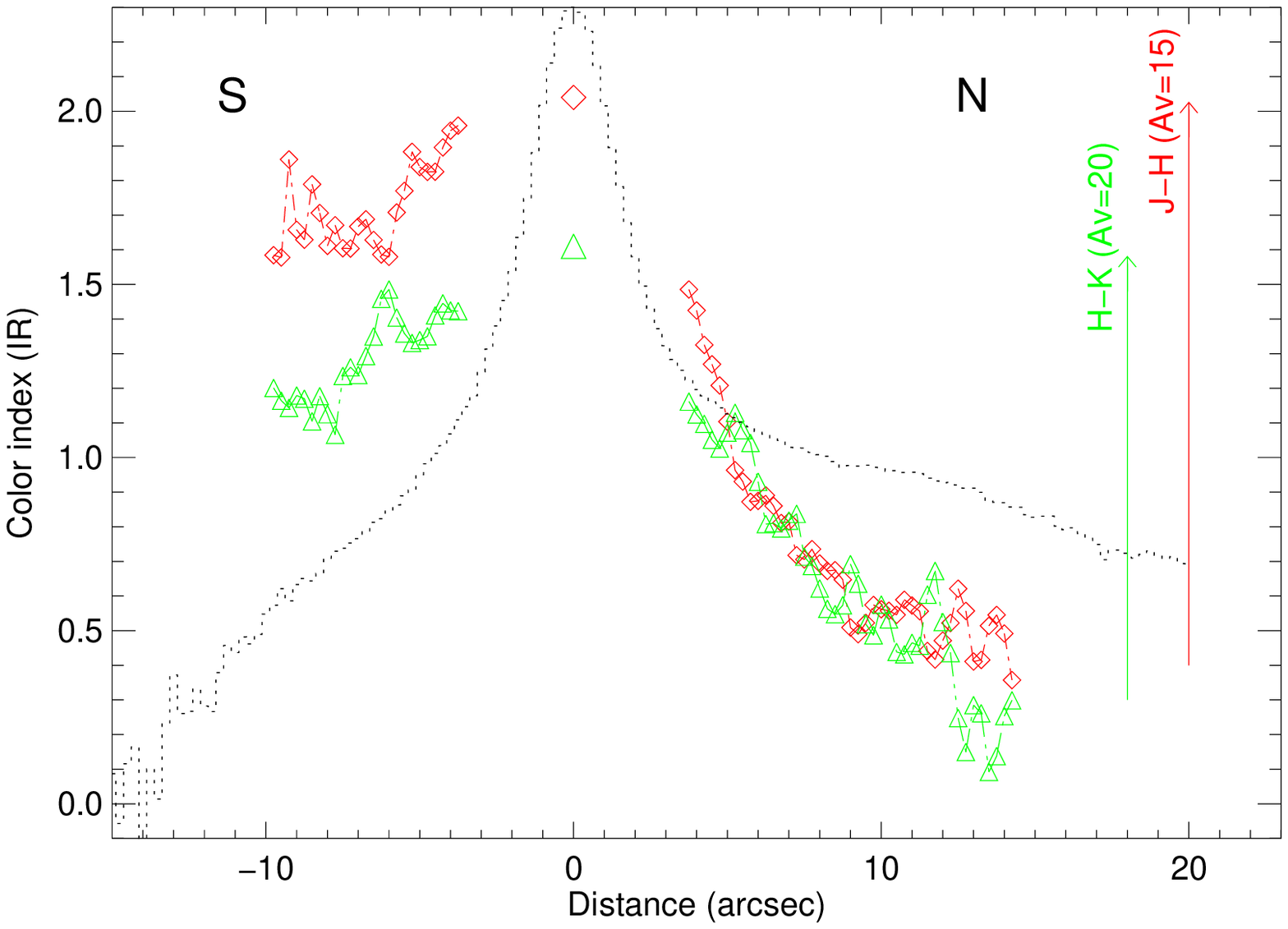}}
\vskip 3mm
\caption{\label{fi:cut_color}
Different color profiles of the nebula along N$-$S, crossing the star.
In the left panel we represent the variation of the $V-R_C$ (stars) and 
$R-I_C$ (crosses) colors,
underlying is the brightness profile as measured in the R band. In the right
panel we represent the variation of the $J-H$ (diamonds) and $H-K_s$ (triangles) 
colors, underlying is the brightness profile as measured in the $J$ band.}
\end{figure*}

In Fig. \ref{fi:cmaps} we show the resulting optical (V$-$R) and near-infrared
(H$-$K) maps. In addition we obtained cross-cuts along the N--S direction
passing through the star (Fig.~\ref{fi:cut_color}).

Some distinguishable regions can be recognised in the $V-R_C$ map:   {\it i --}
the bluest colors  of the nebula are registered towards the North end of the
nebula,    its color is around $\mathrm{V-R_C}\sim 0$ and $\mathrm{R_C-I_C}\sim 0.3$ 
at 60\arcsec\ from the star. Assuming that the light path along this direction
is relatively cleaned of dust, we may observe the colors of the closest 
vicinity of the star, though somewhat modified by scattering and foreground
extinction. These colors are very close to those of  an intermediate  A type
star, indicating the presence of a hot central area probably related to the
outburst of this otherwise cold, late type star \citep{McGehee}.  {\it ii --}
the plume located at a distance between 5 and 10\arcsec\ to the north-west
\citep[position B as defined by][]{Briceno04}  appears redder than the rest of
the nebula,   {\it iii --} a compact region of strong reddening is observed
towards  the direction of V1647~Ori. 

In the $H-K_s$ map the reddest part of the nebula can be seen to  the south of
the illuminating star, outlining a flattened D-shape structure, whose
elongation is nearly perpendicular  to the axis of the optical nebula. The
nebula appears very red along the E--W direction across the star ($J-H \sim
2$ and $H-K_s \sim 1$).  
 When crossing the central part towards
North, the nebula changes  rapidly to bluer colors ($J-H$ and $H-K_s$~$\sim 0.5$),
although the extension of the nebula is smaller than in the visible range.   
The infrared colors are redder at the South compared to the North values at the
same radial distance, indicating a larger extinction along the line of sight in
the South part, or part of emission from the dust envelope. 

The observed color distributions of McNeil's nebula can be interpreted in terms
of a thick circumstellar envelope which intersects the light path from 
the central parts of the system to the scattering grains in the outer envelope.  
The envelope includes  conical outflow cavities perpendicular to the  disk plane,
whose walls scatter the starlight. These cavities had been cleaned up by
previous eruptions, and allow the  light to escape and reach the northernmost
regions of the nebula. The North--South asymmetry is likely caused by the
inclination and the different amount of extinction along our line of sight. A
similar geometry is assumed in  theoretical models of infrared reflection
nebulae around young stellar objects \citep{Lazareff90,Fischer96}.

\subsection{Extended line emission}

\begin{figure*}
\centerline{\includegraphics[width=10cm]{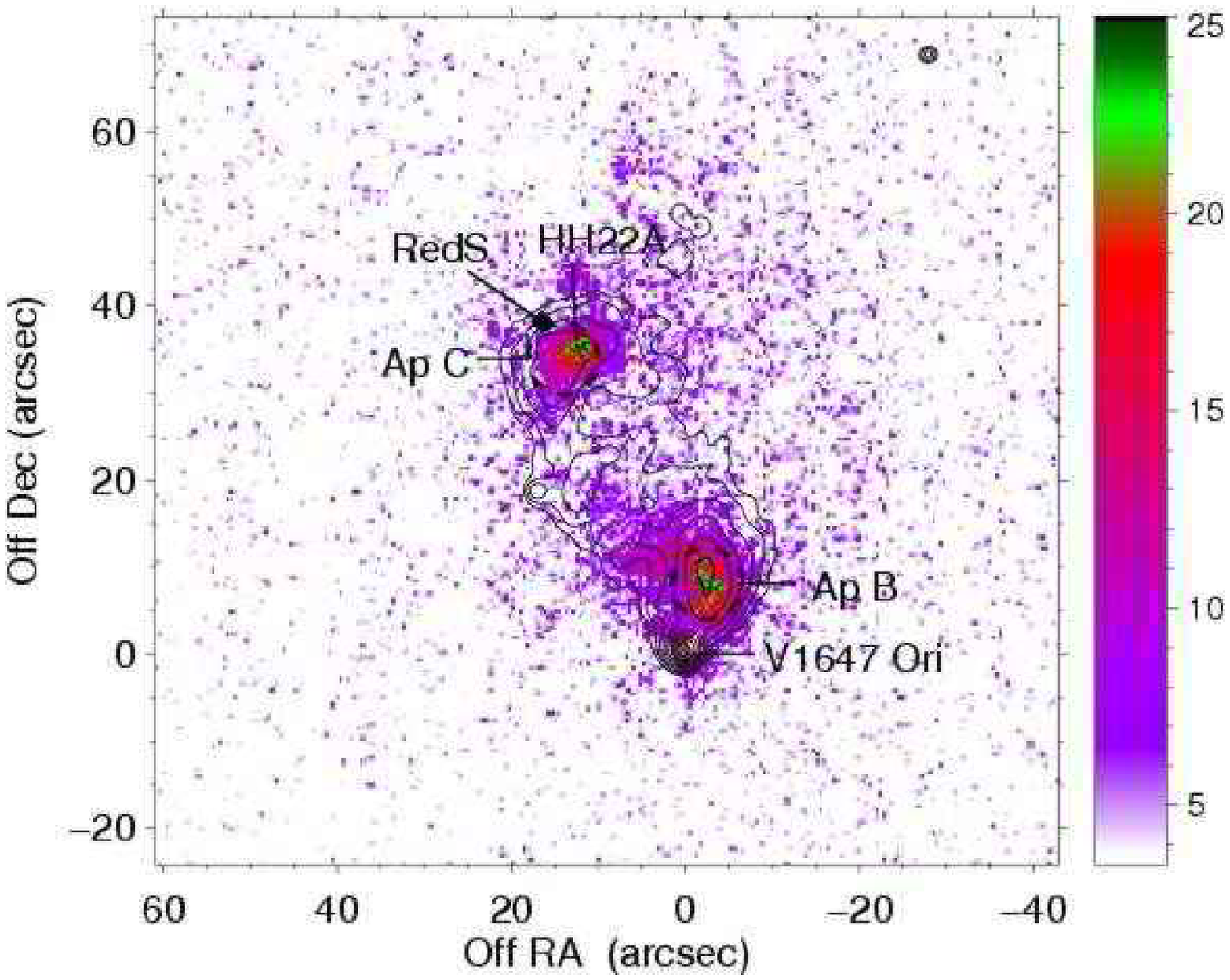}\hspace{5mm}
\includegraphics[width=10cm]{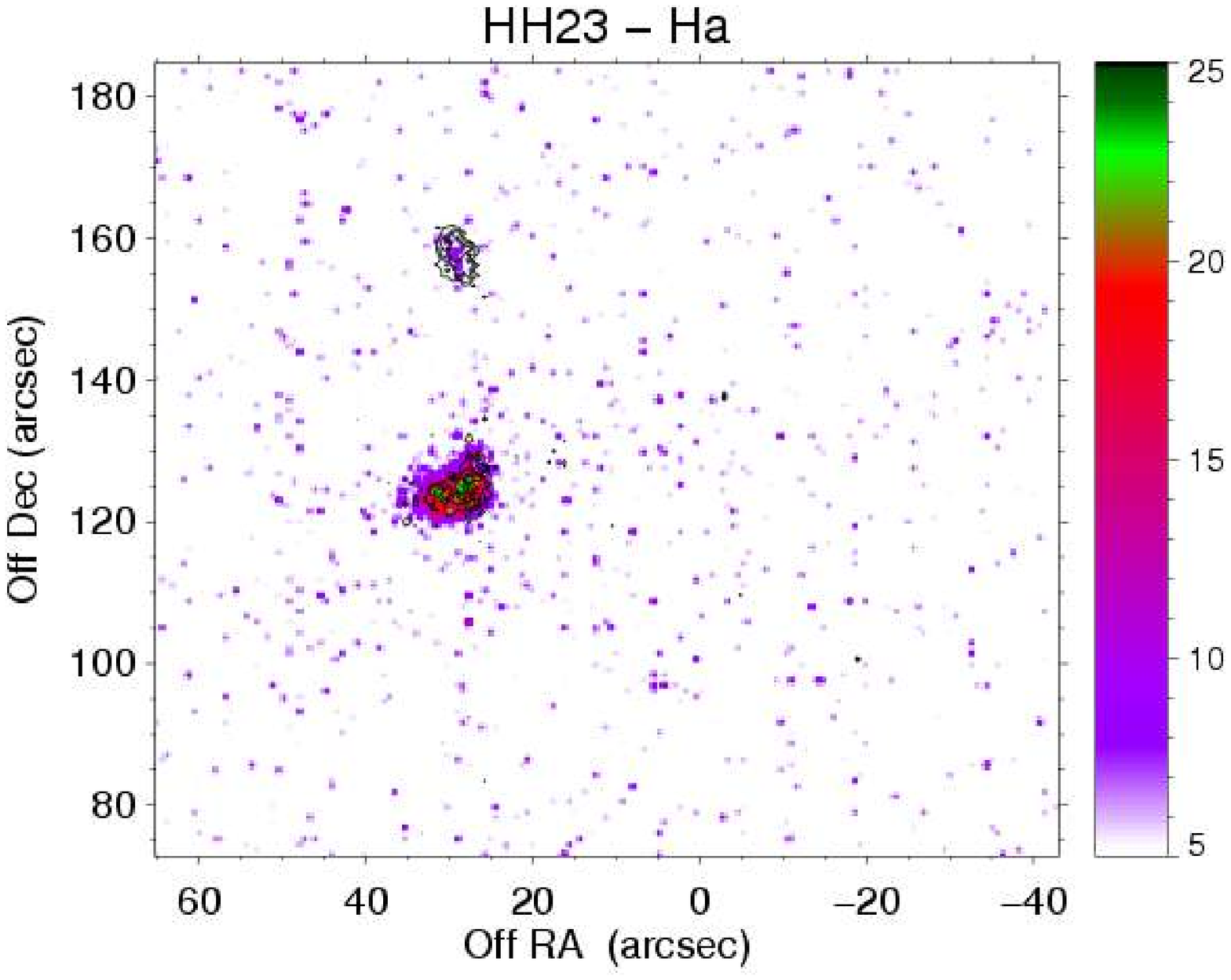}}
\caption{{\it Left panel}: The color scale image corresponds to the 
H$\alpha$ emission after continuum subtraction. The contour lines represent
the emission in the $R_C$ continuum. The filled diamond marks the position 
of the bright infrared source detected
in our $K_s$ band and also by \citet{Muzerolle04} in Spitzer images. Apertures
B and C from \citet{Briceno04} are also indicated.  
{\it Right panel}: 
The color scale image corresponds to the 
H$\alpha$ emission in the region around the Herbig-Haro object HH\,23. 
The contour lines
represent the [SII] emission. Note the bright knot detected in  [SII] 
emission at $\sim 160$\arcsec\ N, which appears very faint in the H$\alpha$. 
The scale is twice that of the left panel.}
\label{Fig_has2}
\end{figure*}

In the left panel of  Fig.~\ref{Fig_has2} we present an emission line H$\alpha$
map, from which  a properly scaled R band image, obtained with the same camera,
had been subtracted.  The morphology observed in the [SII] line map is very
similar,  showing the same two major features around regions B and C.  The
[SII] emission at region C was already reported by \citet{EM} during a
quiescent period of V1647~Ori, and it is very likely associated  to the object
HH22. We detected only knot A in our [SII] images, the other knots B to E are
not seen, very likely because our images are not deep enough.   The emission
at $\sim $~10\arcsec\ North of V1647~Ori (region B) was  not present in the
[SII] map presented by \citet{EM}, and it can be associated with  the outburst
of our target. 

We propose that
the gas responsible for the emission of both H$\alpha$ and [SII] is
ionized by the hot radiation coming from the inner parts of the outburst
region. In this scenario the density of the gas should not exceed  the critical
density of [SII] to avoid collisional de-excitation.  Pure light reflection can
be excluded, partly because the line emitting area  is less extended than 
the region observed  in the adjacent continuum (e.g. $R_C$ or $I_C$ band, see 
Fig.\,\ref{Fig_has2}), and because optical spectra of the central part 
revealed H$\alpha$ but not [SII] emission. 

In the right panel of Fig.~\ref{Fig_has2} we display a region located  North of
McNeil's nebula. The compact object seen at  ${\Delta}Dec \sim 160$\arcsec\ 
corresponds to knot A of the object HH\,23, according to
\citet{EM}. The object  is clearly seen in the [SII]  map, but is faint in
H$\alpha$ emission,  indicative of shock excitation in this region.  The source
V1647~Ori was considered as the source of the HH\,23 outflow by \citet{EM}.  

\subsection{Polarization map}
\label{sc:polariza}

\begin{figure*}
\centerline{\includegraphics[width=10truecm]{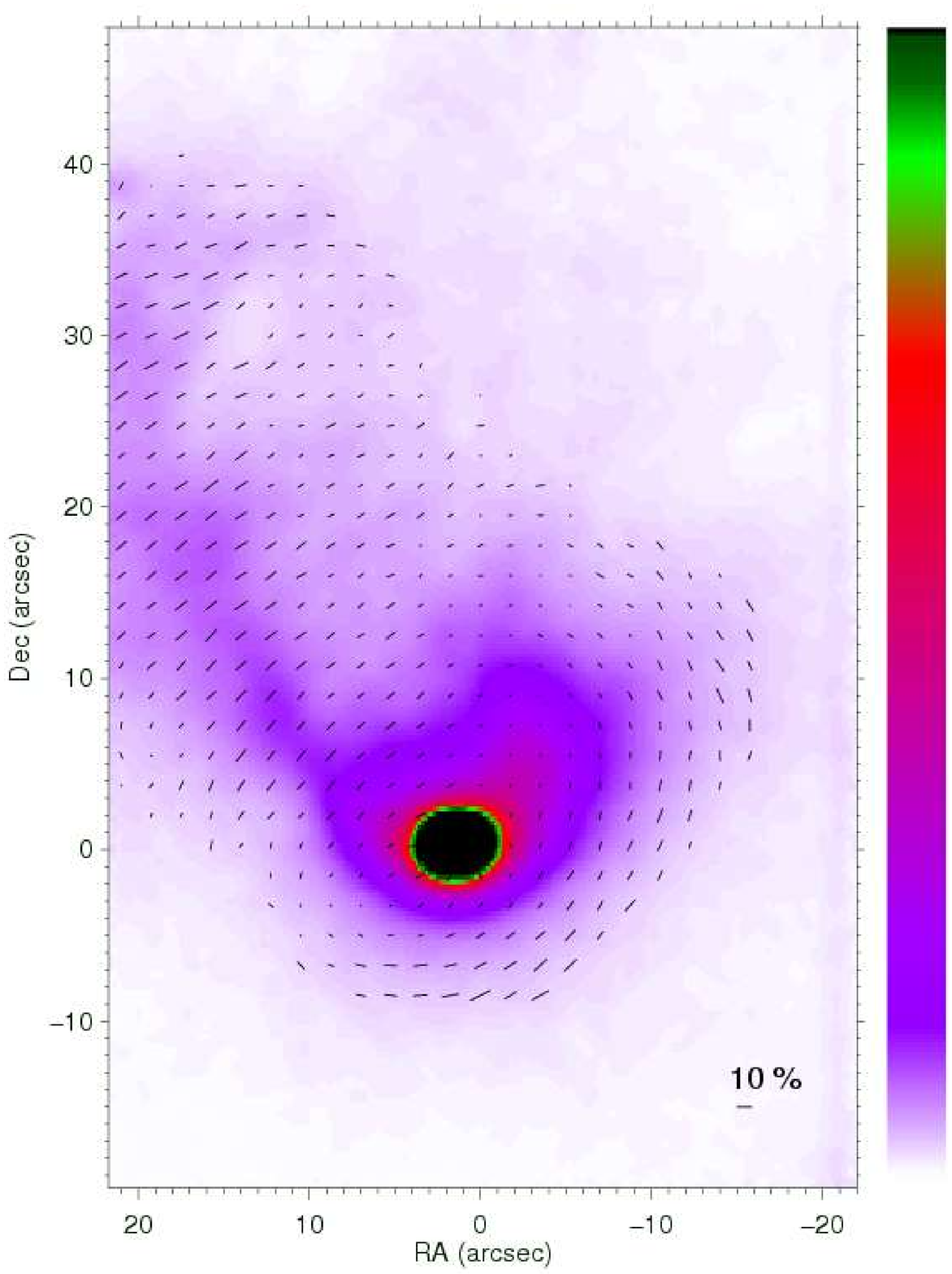}
}
\caption{Polarization map obtained in the J band. The color scaled image
represents the total surface brightness distribution.  
}
\label{fi:jpolorient}
\end{figure*}

Near infrared polarization measurements were obtained using LIRIS in the J band.
The polarization vectors (their orientation and degree of polarization) are
presented in Fig.~\ref{fi:jpolorient}. 

The general  polarization pattern is centrosymmetric around V1647~Ori.  The
largest values of the polarization (slightly above 15\%) are observed along the
North-East rim of the nebula and also, remarkably, to the South--West of the
star. This distribution is consistent with dust scattering taking place in  an
outer envelope surrounding and obscuring the star. Region B, the NW plume close
to the nucleus, shows no significant polarization. There is also a conspicuous
lane of low polarization crossing the star from East to North--West, oriented
almost perpendicularly to the direction of region C.

This pattern of polarization has been observed in other bipolar or cometary
nebulae around young stellar objects \citep{Tamura91}. It is usually attributed
to the dusty disk surrounding the star and it has been successfully simulated by
\citet{Fischer96}. Comparison with their simulations suggests that our
observations are better explained with models which assume dust particles 
of 0.1 to 1~{\micron}~size.  
Among the different accretion disk types our observations seem to be
better reproduced by their SL model, which includes a massive self-gravitating
disk, in the form of an infinite isothermal slab \citep{Lazareff90}. The
polarization towards V1647~Ori (3.2~\% at $\mathrm{PA}= 140^{\circ}$) 
is also consistent both in value and orientation with the simulations  by
\citet{Fischer96}  for the same accretion disk type for the unresolved
polarization. 

\subsection{The very red object near HH\,22} 

Two bright red and compact sources appear in our {\it K\/}$_s$ 
image (Fig. \ref{fi:lirisJoverK}) towards North-East of V1647~Ori. 
The two sources are roughly aligned with the star V1647~Ori. 
They are detected in all our {\it K\/}$_s$ images obtained with LIRIS, even 
during the post--outburst phase when the brightness of V1647~Ori has 
decreased by more than 2 magnitudes. These sources are not detected in our TCS
images due to the poorer sensitive of the camera.   
The closer source is located at the position 
$(\Delta \alpha, \Delta \delta) \simeq (8\arcsec,19\arcsec)$. 
It was not detected in our J images and only barely detected in 
the H images. 
Its estimated magnitudes are $J> 19$, $H > 16, $ and $K_s \simeq 15.0$.
Given the uncertainties in the available data, 
we cannot say more about the nature of this object. 

The farther source is located at an offset of  $(\Delta \alpha, \Delta \delta)
\simeq (16\arcsec,38\arcsec)$.  We estimate the following magnitudes for this
source:  $J> 19$, $H\gtrsim 16, $ and $K_s \simeq 14.5$. 
We have checked the photometry of the source at different epochs, 
showing a variation smaller than 0.3 mags. 
This source has a
corresponding entry in the 2MASS catalog (J05461423-0005261),  whose measured
magnitude is similar to ours  ($\Delta \mathrm{K_s}_s \simeq 0.4$), given the
uncertainties inherent to the faintness of the source.  
Its near infrared colors ($H-K_s \simeq 1.5$ and $J- H >3$) are close to the 
locus of the  classical T Tauri stars \citep{Meyer97}, 
although some reddening is also needed to bring its colors onto this locus.  
\citet{Muzerolle04} detected an infrared source at 4.5 and 
24~\micron\ using the Spitzer Space Telescope. The position of this source, 
as overplotted in Fig.~\ref{fi:lirisJoverK}, shows a perfect 
coincidence with the bright source in our $K_s$ image. In the best 
seeing images this source appears sligthly extended, as already noted
by \citet{Muzerolle04}. 

\citet{Muzerolle04} claimed that this red object is part of  the Herbig-Haro
flow HH\,22, and they propose to classify it as an extreme  embedded class I
source. In Fig. \ref{Fig_has2} we overplotted the position of the red source
using the coordinates given in \citet{Muzerolle04} on our H$\alpha$ emission
line map. It can be seen that the red source is located 3--4\arcsec\ North of 
the H$\alpha$ emission line region which coincides with  the position of knot
A in HH\,22 \citep{EM}.  Thus if the red Spitzer source is considered to be the 
driving source of HH\,22, it should be explained why it is not aligned with the chain
of the knots.  

According to \citet{Briceno04} the location of their region C 
is almost centered on the position of HH22. 
The fact that region C brightened during 
the outburst could be explained by the overlap of two regions:
a dust cloud  which reflects the brightening of the 
star, and the other related to HH22 producing the line emission.

\section{Discussion}

\label{sc:discussion}

The eruptions of young stellar objects are commonly explained as a consequence 
of dramatically enhanced
accretion from the circumstellar disk  onto the star.  Based on this idea,
detailed models for FUors have been developed by \citet{BL} (hereafter BL
model), which were later extended to include trigger mechanism for the
eruption \citep{BL95}, and to take into account  reprocessed light
\citep{tbb}. In this section we first discuss whether the  different types of
observational results on V1647\,Ori, collected during the two years of the
outburst, could be consistently interpreted in the framework of the BL model. 
Then we comment on the question if V1647\,Ori could be considered to be a
member of either the FUor or the EXor group.

\subsection{The outburst mechanism}

\paragraph{Pre-outburst phase.} In the BL model FUor eruptions occur in low
mass pre-main sequence stars  at an early evolutionary phase ($t_{age} \le
10^5$\,yrs). These young objects are still surrounded by an infalling envelope
which feeds material onto an accretion disk. V1647\,Ori seems to be precisely
this type of star: based on pre-outburst measurements, it is a
Class I/Class II object ($t_{age}\approx  10^4$\,yr),  whose flat SED shows
that the object is deeply embedded \citep{AKCs}. 

The rate at which material spirals within the disk towards the star  is a
critical input parameter of the BL model. If accretion exceeds a threshold
value of about $5\times 10^{-7} M_{\odot}$/yr, then most of the matter  would
not fall directly onto the star due to the inefficient outward transportation
of angular momentum. It piles up in the inner disk,  soon  or later leading to
an outburst. In Sect.\,\ref{sec:specevol} we determined the accretion rate of
V1647\,Ori from the Br$\gamma$ line fluxes  for May 2006 when the  star was in
quiescent phase, and obtained a value of  $5\times 10^{-7} M_{\odot}$/yr  (a
similar figure was proposed by \citet{Muzerolle04}). The result agrees with
the  threshold value, indicating that V1647\,Ori just passed the criterion of
the BL model to produce an outburst, but its accretion rate in the disk was
probably the lowest among all known FUors.

\paragraph{The outburst.} According to the BL model, the inflowing matter
piles up in the inner disk at $R_{limit} {\approx}0.25$\,AU, until its column
density -- and thus opacity --  becomes high enough to switch on a thermal
instability. An ionisation front propagates  in the inner disk, and the
emission of the bright ionised region behind  the front causes the increase of
flux at optical and near-infrared wavelengths.   V1647\,Ori brightened by
about 4.5 magnitudes in the $I_{C}$ band, which is a typical value for  FUor
eruptions. One should note, however, that the brightening of V1647\,Ori 
consisted of two effects of comparable  amplitude: an intrinsic brightening
presumably  related to the appearance of  a new hot component in the system
(probably the ionised inner disk); and a dust-clearing event which reduced the
extinction along the line-of-sight \citep{RA, McGehee}.  Thus the amplitude of
the outburst of V1647\,Ori  was probably lower than typical of FUors.  Also,
the bolometric luminosity increased only by a factor of 15, in contrast with
the factor of $\sim$100 of classical FUors.  Concerning the hypothesis of the
BL model that matter piles up at $R_{limit}$, it is interesting to recall that
in Sect.\,\ref{sc:outburst_hist} we speculated that the periodic optical
dimming  could be the consequence of variable extinction due to obscuration by
a dense circumstellar dust clump orbiting the star. The period of the
optical variability of 56 days implies an orbital radius of $R = 0.28\,(M /
M_{\sun})^{1 / 3}\,$AU, which is close to the expected $R_{limit}$ of the
inner disk where material piles up, suggesting that the extinction might be
related to this reservoir of matter.

During the outburst the high temperature in the ionized part of inner disk 
leads to higher  efficiency of angular momentum transport, resulting in
dramatically increased accretion rate onto the star. \cite{Muzerolle04}
determined that the accretion rate increased by a factor of 15. In
Sec.~\ref{sec:specevol} we calculated an accretion rate of $5 \times 10^{-6}
M_{\sun}$/yr from the Br$\gamma$ line luminosity for the peak of the outburst
(close to the estimates of \citet{VCS} and \citet{Gibb06}). These results
suggest that in the eruption of V1647\,Ori both the peak value 
of the accretion rate and its increase with respect to the quiescent phase 
were relatively modest compared to those of other FUors. 

Many observations indicated that a new, hot component appeared in the
V1647\,Ori system during the outburst. This component is probably the inner 
ionised part of the disk, for which the BL model predicts a surface
temperature of 6000-8000\,K. The observations seem to be consistent  with this
expectation: \citet{McGehee} attempted to fit the SED of this new component
with a 7000 K blackbody; \citet{Muzerolle04} assumed a temperature of 6000 K
for the innermost, outbursting part of the disk; and the disk model of
\citet{MIDI} predicted a temperature of 4200 K for the same part. The high
temperature of the inner disk is also supported by the CO excitation diagrams
\citep[2400 K, ][]{Rettig}. It is also consistent with the model of
\citet{BL},  that the region participating in the outburst is limited to 
within a  few AU around the star (see the interferometric measurements by
\citet{MIDI}; and the estimations of \citet{TKKS}). 

The main apparent contradiction to the BL model is related to the timescales.
In their standard self-regulated model, \citet{BL} computed a rise time
of several decades (matching the case of the FUor V1515\,Cyg). In order to
explain shorter initial brightening curves (e.g. V1057\,Cyg, FU\,Ori), they
proposed `triggered eruption' when a nearby companion can temporarily 
increase the accretion rate. V1647\,Ori went into outburst in about 
4 months \citep{Briceno04}, thus it is tempting to speculate on such 
a trigger mechanism; but so far there is no observational signature 
of any companion. One can realize, however, that in the BL model all
timescales depend on the accretion rate in the outer disk: lowering the
accretion rate leads to shorter eruptions. Thus the very low accretion rate
of  V1647\,Ori (practically at the threshold value for producing an outburst)
can be one factor to explain the short initial brightening. Table\,2 in the
paper of \citet{BL} presents rise time estimates for several \.{M}$_{in}$
values. Interpolating within this table suggests, however, that the resulting
timescales are still too long for V1647\,Ori. The other main factor which
determines the timescale of the initial brightening is the viscosity 
parameter in the outburst high state of the inner disk ${\alpha}_h$.
Guessing again from the numbers of Table\,2 of \citet{BL}, we concluded that
a value of ${\alpha}_h = 3\times 10^{-2}$
(rather than their standard value of ${\alpha}_h = 10^{-3}$) would produce
the observed rising timescale of about 4 months.

\paragraph{The plateau phase.}

The initial brightening of V1647\,Ori was followed by a 2-year long slow fading. 
This is the period under which the inner disk depletes and the matter is
becoming neutral again. The length of this period is determined by the same
viscosity parameter as we discussed in relation to the length of the initial
brightening. It is encouraging, that adopting again
${\alpha}_h = 3\times 10^{-2}$ and taking into account the 
low accretion rate of V1647\,Ori, from the numbers in Tab.\,2 of \citet{BL}
we estimated a plateau phase of about two years, consistently with the
observations. 

According to the model of \citet{BL}, the fading is due to the decreasing
accretion rate. Checking two independent indicators of the accretion rate, 
in Sec.~\ref{sec:specevol} we found the somewhat contradictory result 
that the temporal evolution of the flux of the
HI lines indeed shows such a monotonic drop in the accretion rate, while the
CaII\,8542\,\AA{} line fluxes remained constant during the whole outburst. 

The fading rate during this plateau was about 0.04 mag/yr. As mentioned
already in Sect.\,3.1, the fact that the fading rate appears wavelength
independent suggests that no significant temperature variation occurred in the
inner parts of the star-disk system in this period.

\paragraph{Final fading.}
In October 2005, V1647\,Ori started a sudden fading which took approximately 4
months and ended when the object returned to the quiescent, pre-outburst 
state at optical and near-IR wavelengths.
Remarkably, the rate of the brightness change was similar during the final
fading and the initial brightening. This is a natural fact in the BL model,
where the ionization front reverses at the end of the outburst and 
proceeds with the same speed as in the extension phase. We note that such a 
return of a FUor to  the quiescent phase has never been observed before.

\paragraph{Recurrent outbursts?}

Following an eruption, steady accretion from the outer disk slowly
fills up the inner disk at $R_{limit}$, and the system is  prepared 
for a new outburst. The timescale between outbursts is determined by 
the viscosity parameter of the model in the quiescent cold state of the system
${\alpha}_c$. From a rough interpolation in Tab.\,2 of  \citet{BL} one may
conclude that ${\alpha}_c \approx 1.5\times 10^{-3}$ would produce the 40\,yr
period between the last and the present outburst of V1647\,Ori.
This number is larger than the corresponding value in the standard BL model
by a factor of 15.

\paragraph{Dust sublimation/condensation.} Similarly to the initial
brightening, the final fading also consisted of two events: an  intrinsic
fading and an increasing extinction along the line-of-sight. This suggests
that the effects changing the extinction must be reversible. This reversibility is
also supported by the fact that V1647\,Ori had already an outburst in
1966-1967 \citep{Aspin06}. Thus among the possible mechanisms responsible for the
dust-clearing event  sublimation of dust particles at the beginning of the
outburst  is a likely explanation. Is could be a reversible process, since the
increasing extinction observed during the final fading can be due to the
condensation of dust grains (dust condensation effects were already observed
in young eruptive stars, e.g. V1515\,Cyg in 1980, see \citet{v1515cyg}). 

\paragraph{Wind.}
The analysis of spectral lines (Sect.\,\ref{sec:specevol}) suggests 
the presence  of a
wind which was strongest at the peak of the outburst. 
During the plateau phase the P\,Cygni profile of the prominent spectral lines
disappeared, also indicating that the strong wind present at the peak 
of the outburst decreased considerably.
Such a wind is not
included in the model of \citet{BL}, but should be taken into account for a
detailed modelling of V1647\,Ori. The detailed model should also take into
account the interesting finding that the HeI absorption turned into 
emission in the quiescent phase, indicating that the type of the wind 
changed after the end of the outburst.

\paragraph{Conclusions.}

From the comparison of observations of V1647\,Ori with the model 
of \citet{BL} we conclude that the BL model reproduces most observational
results, but in order to match the timescales one order of magnitude 
higher viscosity parameters -- both in the cold and the high states --
have to be assumed. The physical reason behind the increased 
viscosity parameter cannot be deduced from the available data.

\subsection{FUor or EXor?}

In the literature there is an on-going debate on   the
classification of V1647\,Ori, presenting a number of arguments supporting
either the FUor- or the EXor-like nature.  On the one hand, in our previous
paper \citep{AKCs} we argued that the shape of the SED is similar to
FUors, and now we can add that the model of \citet{BL}, developed for explaining
the FUor phenomenon, can model the outburst of V1647\,Ori reasonably well. On
the other hand, the short timescale of the eruption (2 years) and the
recurrent nature argue more for an EXor-type event. EXors, however, are
assumed to be classical T\,Tauri stars, while V1647\,Ori is obviously
in an earlier evolutionary phase than T\,Tau stars. We speculate that 
V1647\,Ori -- together with another recently erupted young stellar object 
OO\,Serpentis \citep{Kospal06} -- might form a new class of young
eruptive stars. Members of this class may be defined by their relatively short 
timescales, recurrent outbursts, modest increase in bolometric luminosity
and accretion rate, and an evolutionary state earlier than that of typical
EXors.  

\section{Summary}
\label{sc:conclu}

Comparison of our optical and near-infrared data obtained in  
2004 February--2006 Sep
on V1647~Ori with published results led to the following new results:
(1) The brightness
of V1647~Ori stayed more than  4\,mag above the pre-outburst level
until October 2005 when it started a  rapid fading. (2) In the high
state we found a periodic component in the optical light curves
with a period of 56 days.  (3) The time delay between the brightness
variations of the star and  a nebular position corresponds to an
angle of $61\degr\pm14\degr$ between the axis  of the nebula and
the line of sight.  (4) The overall appearance of the  infrared and
optical spectra did not change in the period  2004 March--
2005 March, though steady decrease of HI emission line  fluxes
could be observed. We show that the periodic variations of  the
H$\alpha$ equivalent width, suggested by several authors,  could
be caused by variation of the continuum with the 56 days period.
In 2006 May, in the quiescent phase, the HeI 1.083 $\mu$m line was
in  emission, contrary to its deep blueshifted absorption observed
during the outburst. (5) The $J-H$ and $H-K_s$ color maps  of the
infrared nebula reveal an envelope around the star whose largest 
extension is about 18\arcsec (0.03\,pc). (6) The color distribution
of the infrared  nebula suggests reddening of the scattered light
inside a thick circumstellar disk. (7) Comparison of the {\it K\/}$_s$  
and H$\alpha$
images of McNeil's Nebula shows that HH\,22A, the Spitzer infrared source 
and the bright clump C
of the nebula may be unrelated objects. (8) We show that the observed
properties of V1647~Ori could be interpreted  in the framework of
the thermal instability models of \citet{BL95}. (9) We speculate
that V1647~Ori might belong to a new class of young eruptive
stars, defined by relatively short  timescales, recurrent
outbursts, modest increase in bolometric luminosity and accretion
rate, and an evolutionary state earlier than that of typical EXors.  

\acknowledgements

Financial support from the Hungarian OTKA grants K62304, 
T42509 and T49082 is acknowledged.
AM, JAP, MCLl and MJVN acknowledge support from grant AYA 2001-1658, financed by 
the Spanish Direcci\'on General de Investigaci\'on. 
Many thanks are due to Maria J. Arevalo and Gabriel G\'omez, 
Jos\'e A. Caballero and Fernanda Artigue 
for helping with the data collection at the IAC-80 and TCS telescopes. 
We also thank Nicola Caon for his continuous help with all IRAF related questions.
This project has been supported by the Australian Research Council. 
LLK is supported by a University of Sydney Postdoctoral Research Fellowship.

\appendix
\section{The 2MASS source J05461162--0006279}

The object located at about 30\arcsec\ West and 30\arcsec\ S of V1647~Ori 
(Fig.\,\ref{fi:lirisJoverK})
can be identified with the 2MASS source J05461162$-$0006279. The 
source is very red as shown by its appearance as a remarkably bright 
object in Spitzer images at 4.5 and 24 \micron\ \citep{Muzerolle04}.
It was also detected by XMM \citep{Grosso}, however it remains undetected 
in optical images.
Thanks to the excellent seeing ($0\farcs65$) during the 
night 2004  Nov 4 we could resolve this 
target into two components separated by $1\farcs2$, oriented along 
the E--W axis (see Fig. \ref{fi:lirisJoverK}). 
The magnitude of each component was obtained by performing
PSF photometry using the daophot package in IRAF. The primary star is
the Western component, and has  
J$=14.14$ and K$_s =11.43$; 
and the secondary star has J$=15.72$  and K$_s=13.79$. 
The color index J$-$K$_s$ is 2.7 and 1.9 for the
primary and secondary stars, respectively. These color indices are redder
than those of normal M stars according to \citet{Bessel88}, suggesting 
the possibility
that this double system is also embedded in the Orion\,B molecular cloud.


\end{document}